\documentclass[structabstract]{aa}  
\usepackage{graphicx}
\usepackage{natbib}
\usepackage{lscape}
\usepackage{longtable}

\usepackage{txfonts}
\def\macc   {$\dot{M}_{\rm acc}$}
\def\lacc   {$L_{\rm acc}$}

\def\msun {$M_{\odot}$}
\def\lsun {$L_{\odot}$}
\def\lstar {$L_*$}
\def\rin {$R_{\rm in}$}
\newcommand{\degree}{\ensuremath{^\circ}}

\begin{document}
   \title{On the gas content of transitional disks: \\a VLT/X-Shooter study of accretion and winds\thanks{This work is based on observations made with ESO Telescopes at the La Silla Paranal Observatory under programme ID  089.C-0840 and 090.C-0050, and on data obtained from the ESO Science Archive Facility observed under programme ID 084.C-1095, 085.C-0764, 085.C-0876, 288.C-5013, and 089.C-0143. }} 

   \author{C.~F.~Manara\inst{1}, L.~Testi\inst{1,2,3}, A.~Natta\inst{2,4}, G. Rosotti\inst{5,6}, M. Benisty\inst{7}, B. Ercolano\inst{5,3}, L.~Ricci\inst{8}}

   \institute{European Southern Observatory, Karl Schwarzschild Str. 2, 85748 Garching, Germany\\
              \email{cmanara@eso.org}
         \and
             INAF - Osservatorio Astrofisico di Arcetri, Largo E.Fermi 5, I-50125 Firenze, Italy
	\and
		Excellence Cluster Universe, Boltzmannstr. 2, D-85748 Garching bei M{\"u}nchen, Germany
	\and
	School of Cosmic Physics, Dublin Institute for Advanced Studies, 31 Fitzwilliam Place, Dublin 2, Ireland
	\and
		Universit{\"a}ts-Sternwarte M{\"u}nchen, Scheinerstr. 1, 81679 M{\"u}nchen, Germany
	\and
		Max Planck Institut f{\"u}r Extraterrestrische Physik, Giessenbachstrasse 1, 85748 Garching bei M{\"u}nchen, Germany 
	\and
		 Institut de Plan{\`e}tologie et Astrophysique Grenoble, 414 rue de la Piscine, 38400 St-Martin d'H{\`e}res, France  
	\and
		California Institute of Technology, 1200 East California Boulervard, 91125 Pasadena, CA, USA
		}

   \date{Received December 22, 2013; accepted June 4, 2014}

 
  \abstract
   { Transitional disks are thought to be a late evolutionary stage of protoplanetary disks whose inner regions have been depleted of dust. The mechanism responsible for this depletion is still under debate. To constrain the various models it is mandatory to have a good understanding of the properties of the gas content in the inner part of the disk. }
   {Using X-Shooter broad band - UV to NIR - medium resolution spectroscopy we derive the stellar, accretion, and wind properties of a sample of 22 transitional disks. The analysis of these properties allows us to put strong constraints on the gas content in a region very close to the star ($\lesssim$ 0.2 AU) which is not accessible with any other observational technique. }
   {We fit the spectra with a self-consistent procedure to derive simultaneously spectral type, extinction, and accretion properties of the targets. From the continuum excess at near-infrared wavelength we distinguish whether our targets have dust free inner holes. Analyzing forbidden emission lines we derive the wind properties of the targets. We then compare our findings to results for classical TTauri stars.}
   {The accretion rates and wind properties of 80\% of the transitional disks in our sample, which is strongly biased towards stongly accreting objects, are comparable to those of classical TTauri stars. Thus, there are (at least) some transitional disks with accretion properties compatible with those of classical TTauri stars, irrespective of the size of the dust inner hole. Only in 2 cases the mass accretion rates are much lower, while the wind properties remain similar. We do not see any strong trend of the mass accretion rates with the size of the dust depleted cavity, nor with the presence of a dusty optically thick disk very close to the star. These results suggest that, close to the central star, there is a gas rich inner disk with density similar to that of classical TTauri stars disks.}
  {The sample analyzed here suggests that, at least for some objects, the process responsible of the inner disk clearing should allow for a transfer of gas from the outer disk to the inner region. This should proceed at a rate that does not depend  on the physical mechanism producing the gap seen in the dust emission and results in a gas density in the inner disk similar to that of unperturbed disks around stars of similar mass. }


 \authorrunning{Manara et al.}
\titlerunning{X-Shooter study of accretion and winds in Transitional Disks}
\maketitle
%

\section{Introduction}

At the beginning of their evolution, protoplanetary disks surrounding forming stars appear as a continuous distribution of gas and small dust particles. For the first few Myrs they evolve by viscous accretion \citep{Hartmann98}, and in the meantime grain growth and planet formation take place. The observations show that, in a relatively small fraction of objects ($\sim$10 \%, e.g., \citealt{EspaillatPPVI}), a significant change in the disk morphology is detected: a dust-depleted region in the inner part of the disk appears in mm-interferometry observations \citep[e.g.][]{Andrews11} and/or as a dip in the mid-IR spectral energy distribution \citep[SED; e.g.,][]{Merin10}. This can be either a {\it hole} - absence of gas from the dust sublimation radius out to some much larger radius - or a {\it gap} - absence of gas in a relatively narrow region, like a ring. These disks are known as transitional disks \citep[TDs; e.g.,][]{Calvet05,EspaillatPPVI}. In some cases an excess emission is detected at near-infrared wavelengths; this emission comes from a small annulus of warm dust close to the star \citep[e.g.,][]{Benisty10}. These objects are commonly referred to as pre-transitional disks \citep[PTDs; e.g.,][]{Espaillat10,EspaillatPPVI}.

Different processes have been proposed to explain the formation of these gaps/holes. The most plausible mechanisms so far are photoevaporation, grain growth, or planet formation \citep[e.g.,][]{EspaillatPPVI}. Still, none of these processes alone has been shown to be sufficient to explain all observations. Grain growth models explain the observed IR SEDs of TDs, but are unable to reproduce mm-observations \citep{Birnstiel12}. The detection of large mass accretion rates (\macc) in several TDs with large inner hole sizes is at odds with photoevaporative model predictions, which expect a very fast depletion of the gas mass reservoir for accretion by the inner disk onto the star \citep[e.g.][]{Owen11,Owen12}. Moreover, planet formation models, which need to include multiple accreting planets to perturb sufficiently the inner disk surface density, still cannot explain TDs with large inner hole sizes and high mass accretion rates \citep{Zhu11}. At the same time, the presence of a planet could explain the radial and azimuthal distribution of mm-sized grain particles in TDs \citep{Pinilla12}. New attempts are being made to include various processes in one model. As an example, \citet{Rosotti13} combined photoevaporation and planet formation, but they were still not able to explain the observed accretion properties of many TDs. Similar constraints arise from observations of winds in TDs. Analysis of forbidden line emission show that winds are emitted from the innermost region of the disk in various objects \citep{AlexanderPPVI}, and these observations suggest that photoevaporation could play a role in the clearing of disks. At the same time, it is not yet clear what is the origin and what are the properties of these winds, and more studies are needed to have a clearer understanding of this aspect.

In this context, the combination of inner hole size, \macc, and wind properties is a powerful observational diagnostic of disk evolution models. In particular, \macc \ and the wind properties allow us to place strong constraint on the gaseous content of the innermost region of these disks which can be compared with the models. As explained by magnetospheric accretion models \citep[e.g.][]{Hartmann98}, the process of accretion is related to the gaseous content of the innermost region of the disk at radii $\lesssim$ 0.2 AU. Similarly, forbidden lines are emitted in regions in the disk as close as $\sim$ 0.2 AU from the central star. The measurements of  \macc \ for TDs available in the literature are mostly based on secondary indicators (such as the 10\% H$\alpha$ width) and have been obtained using different non-homogeneous techniques. In many cases these values are highly uncertain and, therefore, not reliable. At the same time, very few data on the wind properties for TDs are available. In order to remedy these deficiencies we have collected a sample of 22 spectra of TDs with the ESO VLT/X-Shooter spectrograph. We aim at deriving with an highly reliable method the stellar and accretion properties of these objects and to study simultaneously their wind properties from optical forbidden lines. 

The analysis of this sample of TDs that we present here is focused firstly on deriving the values of \macc \ for these objects in order to verify the reliability of those reported in the literature. We use a very detailed and self-consistent analysis to derive accretion rates and simultaneously the spectral types and stellar properties of the objects from the fit of the whole spectrum from UV to NIR. In particular, we want to check the high values of \macc \ for objects with large inner hole sizes, that cannot be explained by current models. At the same time, we want to determine whether there is any dependence of the accretion properties of TDs with their disk morphology, in particular testing possible correlation with the inner hole size. Moreover, we investigate the differences and similarities in accretion and wind properties of TDs with respect to classical TTauri stars. Finally, we put constraint on the properties of the gaseous innermost regions of the disk in these objects from the derived values of \macc \ and from the wind properties.

The paper is organized as follows. In Sect.~\ref{sect::obs} we present the observations, the data reduction procedure, and the properties of the targets in our sample. In Sect.~\ref{sect::method} we briefly describe the method used to derive the stellar and accretion properties of the objects, and we report the derived values. Then, in Sect.~\ref{sect::wind} we derive the wind properties of our targets. In Sect.~\ref{sect::disc} we discuss our results and describe the additional data from the literature collected to derive our conclusions, which we summarize in Sect.~\ref{sect::concl}.


\section{Observations}
\label{sect::obs}

All the observations included in this work have been obtained with the ESO/VLT X-Shooter spectrograph. This medium resolution and high-sensitivity instrument covers simultaneously the wavelength range between $\sim$300 nm and $\sim$ 2500 nm, dividing the spectrum in three arms, namely the UVB arm in the region $\lambda\lambda\sim$ 300-560 nm, the VIS arm between $\lambda\lambda\sim$ 560-1020 nm, and the NIR arm from $\lambda\sim$ 1020 nm to $\lambda\sim$ 2500 nm \citep{Vernet11}. In the following, we present the properties of the sample, the details of the observations, that are reported also in Table~\ref{tab::obs}-\ref{tab::Cl3}, and the data reduction procedure.

\subsection{Sample description}\label{sect::sample}

The first criterion used to select the objects in our sample was to include all the targets with known inner hole sizes (\rin) larger than $\sim$ 20 AU and large \macc \ ($\gtrsim 10^{-9} M_\odot$/yr). These were selected mainly from the sample of \citet{Andrews11}, where the value of \rin \ has been measured using resolved mm-interferometry observations. From this sample we selected the 8 objects with spectral type later than G2. For two of these objects (LkCa15, ISO-Oph 196) the X-Shooter spectra were available in the ESO archive, while we observed the remaining six targets (LkH$\alpha$330, DM Tau, GM Aur, RX J1615-3255, SR21, and DoAr 44) during our programs (see Table~\ref{tab::obs}). We added to these objects also four TDs for which \rin \ was derived from IR SED fitting by \citet{Merin10} and \citet{Kim09}, namely SZ Cha, CS Cha, Sz 84, and Ser 34. Only for the latter the spectrum was not available in the ESO archive.

Then, we included some objects with smaller inner hole sizes and different values of \macc, both as high as the object with large \rin \ and smaller than those. In particular, we included five targets whose spectra were not available in the ESO archive and with \rin $\lesssim$ 15 AU and as small as 1 AU, namely Oph22, Oph24, and Ser29 from the sample of \citet{Merin10}, RX J1842.9 and RX J1852.3 from \citet{Hughes10}. Finally, we collected all the spectra of TDs classified by \citet{Kim09} available in the ESO archive (four objects, CHXR22E, Sz 18, Sz 27, and Sz 45) and the spectrum of TWHya, whose \rin \ has been measured with resolved mm-observations by \citet{Hughes07}. In total the sample analyzed here comprises 22 objects. 

For 9 of the objects analyzed here the value of \rin \ has been directly determined from resolved mm-interferometry observations \citep{Hughes07,Andrews11}, while for the remaining 13 targets the classification as TD and the size of the inner hole has been determined from IR SED fitting \citep{Kim09,Merin10,Hughes10,Espaillat13}. The list of targets, their distances, and the values of \rin \ available in the literature are reported in the first three columns of Table~\ref{tab::XS}. The objects are located in different star forming regions (Perseus, Taurus, Chameleon, TW Hydrae, Lupus, $\rho$-Ophiucus, Serpens, Corona Australis) and have values of \rin \ between $\sim$1 to $\sim$ 70 AU, being representative of the whole range of measured values of \rin. When both values of \rin \ obtained using IR SED fitting and mm-interferometry resolved observations are available we adopt in the analysis the mm-interferometry result. More information on individual objects in the sample are reported in Appendix~\ref{app::peculiar}.

Even if the sample is containing objects of different TD morphologies, such as various inner hole sizes, this is not statistically complete and it is in general biased toward accreting TDs. As explained before, our selection criteria were aimed at observing TDs with already known and high accretion rates, thus our own observations represent a biased sample. On the other hand, the targets collected in the literature were in some cases selected with different criteria that could mitigate our biases. Unfortunately, it is not possible to estimate completely the bias in its selection.

\subsubsection{Class~III properties}
\label{sec::cl3}

Here we present the properties of three non-accreting (Class~III) YSOs, that we use as photospheric templates in our analysis (see Sect.~\ref{sect::method}) to enlarge the available sample of Class~III YSOs observed with X-Shooter and presented in \citet{Manara13a}. We follow the same procedure as in \citet{Manara13a} to derive their spectral types and stellar properties, which are reported in Table~\ref{tab::Cl3}.  

The YSO IC348-127 has spectral type (SpT) G4 \citep{Luhman98,Luhman03} and has been classified as Class~III from \citet{Lada06} using Spitzer photometry. This classification, with values of extinction $A_V$ $\sim$ 6 mag, has been confirmed by \citet{Cieza07} and \citet{Dahm08}. We confirm the spectral classification and the extinction, and we derive $L_*$ = 12.9$\pm$5.9 \lsun \ for this object. Due to the very high $A_V$, the spectrum of this target at $\lambda\lesssim$350 nm is very noisy.

The second Class~III YSO included in this work is T21, which has been classified as a Class~III YSO with SpT G5 by \citet[][and references therein]{Manoj11}. The typical reddening law of the Chameleon~I region in which this object is located is not well constrained \citep{Luhman08}, and could be described using values of $R_V$ \citep{Cardelli} up to 5.5. By comparison of the dereddened spectrum with a blackbody at $T$=5770 K, which is the typical $T_{\rm eff}$ of a star with SpT G5, we obtain that the extinction towards this object is better represented using $R_V$ = 3.1 and $A_V$ = 3.2 mag. Adopting these values, the derived luminosity of the target is $L_*$ = 18.5$\pm$8.5 \lsun.

Finally, we include in the sample CrA75, which has been classified as a Class~III YSO with SpT K2 from \citet[][and references therein]{Forbrich07}. This has been later confirmed by \citet{Peterson11} and \citet{Currie11}, who suggested that the correct values of extinction for this object is $A_V$=1.5, assuming $R_V$=5.5, representative of objects in the Corona Australis region \citep{Peterson11,Chapman09}. With these parameters we derive for CrA75 $L_*$ = 0.4$\pm$0.2 \lsun.

The objects whose properties have been described in this section expand the coverage in SpT of our library of photospheric template. This, however, remains incomplete. In particular, the objects presented in \citet{Manara13a} have an almost uniform coverage in the SpT range from K5 to M6.5. The three objects presented here, instead, do not cover entirely the range of SpT from G3 to K5. This incompleteness of photospheric templates in this range will be considered in the analysis.

\subsection{Observational strategy}
As explained before, we have included in the analysis both new observations and archival data. In the following, we describe separately our observational strategy and the settings used in the archival observations.

\subsubsection{New observations}
New observations with the ESO/VLT X-Shooter spectrograph have been carried out in service mode between April and November 2012  (ESO Pr.Id. 089.C-0840 and 090.0050, PI Manara). The targets have been observed in ABBA slit-nodding mode to achieve the best possible sky subtraction also in the NIR arm. The objects have been observed using different slit widths in the UVB arm. For the brightest objects the slit 0.5x11\arcsec, which leads to the highest spectral resolution in this arm (R=9100), has been adopted, while for the fainter objects - namely Oph22, Oph24, Ser29, and Ser34 - we have used the 1.0x11\arcsec slit, which leads to a lower resolution (R=5100) but allows to achieve an higher S/N. In the VIS and NIR arms the 0.4x11\arcsec slit has been adopted for all the targets. This slit width leads to the highest possible resolution (R= 17400, 10500 in the VIS and NIR arms, respectively) and to enough S/N in the spectra. The readout mode used was in all cases ``100,1x1,hg". To obtain a better flux calibration, we observed the targets of Pr.Id. 090.0050 with the large slit (5.0x11\arcsec) immediately after the exposure with the narrow slit. With the large slit we obtain spectra with a lower resolution but, at the same time, we avoid slit losses and we get a reliable flux calibration for this spectrum, which is then used to calibrate the narrow slit one (see Sect.~\ref{sec::red}). The names of the targets, their coordinates, observing date, and exposure times of the observations are summarized in Table~\ref{tab::obs}.

We observed in our programs also two Class~III YSOs (see Sect.~\ref{sec::cl3} for details). We observed CrA 75 using the narrower slits in each arm - 0.5x11\arcsec \ in the UVB arm, 0.4x11\arcsec \ in the VIS and NIR arms - to obtain the highest possible spectral resolution. For IC348-127, we adopted the slit widths 1.6x11\arcsec, 1.5x11\arcsec, and 1.2x11\arcsec \ in the UVB, VIS, and NIR arms, respectively. This was done in order to achieve enough S/N in the available observing time. We recap in Table~\ref{tab::Cl3} these information.

\subsubsection{Archival data}

The data included in our analysis collected from the ESO archive have been obtained using different observational strategies. The observational data are presented here and summarized in Table~\ref{tab::obs} and \ref{tab::Cl3}.

The transitional disk Sz 84 has been observed during the INAF GTO time in Pr.Id. 089.C-0143 (PI Alcal\`{a}). The adopted slits width for this object were 1.0x11\arcsec in the UVB arm and 0.9x11\arcsec in the VIS and NIR arms. More details on the observing procedure and on the data reduction for this targets are given in \citet{Alcala14}. 

The target ISO-Oph 196 was observed for the program Pr.Id. 085.C-0876 (PI Testi) using the same settings as the one adopted in our observations. We include in our analysis two targets - namely DoAr44 and TW Hya - from the program Pr.Id. 085.C-0764 (PI Guenther). Both targets have been observed with the narrow slits. For all these observations the readout mode used was ``100,1x1,hg", as in our programs, and for each object 4 exposures in the ABBA slit-nodding mode were taken. 

We consider in our work also seven objects from the program Pr.Id. 084.C-1095 (PI Herczeg). Six of those - namely CS Cha, CHXR22E, Sz18, Sz27, Sz45, and Sz Cha - are TDs, while T21 is a Class~III YSOs. These targets have been observed both with a narrow-slit setting (slit widths 1.0x11\arcsec in the UVB and 0.4x11\arcsec \ in the VIS and NIR arms) and with the large slit to have a better flux calibration of the spectra. The narrow-slit observations have been carried out with the ``400,1x2,lg" mode with a AB slit-nodding mode. The large-slit exposures have been obtained in stare mode. 

The data for the TD LkCa15 have been obtained in the program Pr.Id. 288.C-5013 (PI Huelamo). We use in our analysis only 4 exposures obtained in one epoch (2011-12-01), which correspond to an entire ABBA slit-nodding cycle. Observations have been made using the 0.8x11\arcsec \ slit in the UVB arm, the 0.7x11\arcsec \ one in the VIS arm, and the 0.9x11\arcsec \ slit in the NIR arm. The readout mode used was ``100,1x1,hg". Using only 4 frames we obtain a spectrum with enough S/N for our purpose.

\subsection{Data reduction}
\label{sec::red}
Data reduction has been carried out using the version 1.3.7 of the X-Shooter pipeline \citep{Modigliani}, run through the {\it EsoRex} tool. The spectra were reduced independently for the three spectrograph arms. The pipeline takes into account, together with the standard reduction steps (i.e. bias or dark subtraction, flat fielding, spectrum extraction, wavelength calibration, and sky subtraction), also the flexure compensation and the instrumental profile. We checked with particular care the flux calibration and telluric removal of the spectra. 

Telluric removal has been performed using the standard telluric spectra that have been provided as part of the standard X-Shooter calibration plan on each night of observations. Spectra of telluric standard stars observed at similar airmasses right before or after the target have been selected. The correction has been accomplished using the IRAF\footnote{IRAF is distributed by National Optical Astronomy Observatories, which is operated by the Association of Universities for Research in Astronomy, Inc., under cooperative agreement with the National Science Foundation.} task {\it telluric} adopting the same procedure for telluric normalization in the VIS and for response-function preparation in the NIR as explained by \citet{Alcala14}.

Flux calibration has been carried out within the pipeline. Then, for the targets where only narrow-slit observations were available, we checked the flux-calibrated pipeline products comparing them with the available photometry to quantify slit losses. These spectra were then rescaled to the photometric data, and a final check was performed to verify a correct conjunctions between the three arms. The overall final agreement is very good. On the other hand, in the cases where observations with the large slit were available, we first checked that the flux-calibration of the spectra obtained with this slit were compatible with the available photometry. Then, we rescaled the narrow-slit spectra to the large-slit flux-calibrated ones, thus achieving the best possible flux calibration. Also in this case the final products have very good conjunctions between the arms.

\begin{landscape}
\begin{table}[!]
\caption{\label{tab::obs}Transitional disks observing log}
\begin{tabular}{ll | cc | c | ccc | c }
\hline\hline

Name & Other names & RA (2000) & DEC (2000) & OBS. DATE & \multicolumn{3}{c}{T$_{\rm exp}$ (sec)} & Pr.Id (PI)\\
\hbox{} & \hbox{} & h m s & \degree \ \arcmin \ \arcsec & YY-MM-DD & UVB & VIS & NIR & \hbox{}\\

\hline

LkH$\alpha$330 & ... & 03 45 48.29 & 32 24 11.9 & 2012-12-18 & 4x150 & 4x150 & 4x150 & 090.C-0050 (Manara)  \\ 
DM Tau & ... & 04 33 48.74 & 18 10 09.7 & 2012-11-23 & 4x300 & 4x300 & 4x300 & 090.C-0050 (Manara) \\ 
LkCa15 & ... & 04 39 17.79 & 22 21 03.4 & 2012-03-06 & 4x200 & 4x220 & 4x200 & 288.C-5013 (Huelamo)  \\ 
GM Aur & ... & 04 55 10.98 & 30 21 59.3 & 2012-11-23 & 4x300 & 4x300 & 4x300 & 090.C-0050 (Manara)  \\ 
SZ Cha & Ass Cha T 2-6 &  10 58 16.77 & -77 17 17.0 & 2010-01-18 & 4x140 & 4x150 & 4x150 & 084.C-1095 (Herczeg) \\ 
TW Hya & ... &  11 01 51.91 & -34 42 17.0 & 2010-05-03 & 4x150 & 4x60 & 4x100 & 085.C-0764 (Guenther)  \\ 
CS Cha & ISO ChaI 3 & 11 02 24.91 & -77 33 35.7 & 2010-01-18 & 2x55 & 2x60 & 2x60 & 084.C-1095 (Herczeg)  \\ 
CHXR22E	& ... &  11 07 13.30 & -77 43 49.9 & 	2010-01-19 & 2x280 & 2x300 & 2x300 & 084.C-1095 (Herczeg) \\ 
Sz18 & Ass Cha T 2-25 &  11 07 19.15 & -76 03 04.8 & 2010-01-17 & 2x60 & 2x53 & 2x60 & 084.C-1095 (Herczeg)  \\ 
Sz27	& Ass Cha T 2-35 &  11 08 39.05 & -77 16 04.2 & 	2010-01-18 & 2x410 & 2x420 & 2x420 & 084.C-1095 (Herczeg)  \\ 
Sz45	& Ass Cha T 2-56 &  11 17 37.00 & -77 04 38.1  & 	2010-01-17 & 2x45 & 2x40 & 2x45 & 084.C-1095 (Herczeg)  \\ 
Sz84 & ... & 15 58 02.53 & -37 36 02.7 & 2012-04-18 & 2x350 & 2x300 & 2x115 & 089.C-0143 (Alcal\`{a}) \\
RX J1615-3255 & 2MASS J16152023-3255051 & 16 15 20.23 & -32 55 05.1 & 2012-05-21 & 4x300 & 4x300 & 4x300 & 089.C-0840 (Manara) \\
Oph22 & SSTc2d J162245.4-243124 & 16 22 45.40 & -24 31 23.7 & 2012-04-25 & 4x600 & 4x600 & 4x600 & 089.C-0840 (Manara)  \\ 
Oph24 & SSTc2d J162506.9-235050 & 16 25 06.91 & -23 50 50.3 & 2012-06-07 & 4x600 & 4x600 & 4x600 & 089.C-0840 (Manara)\\ 
SR 21 & EM*SR21A, ISO-Oph 110 & 16 27 10.28 & -24 19 12.7 & 2012-04-14 & 4x300 & 4x300 & 4x300 & 089.C-0840 (Manara)\\ 
ISO-Oph 196 & WSB 60 & 16 28 16.51 & -24 36 57.9 & 2010-07-28 & 4x750 & 4x750 & 4x750 & 085.C-0876 (Testi)  \\ 
DoAr 44 & ... &  16 31 33.46 & -24 27 37.2 & 2010-05-03 & 4x300 & 4x140 & 4x150 & 085.C-0764 (Guenther)  \\ 
Ser29 & SSTc2d J182911.5+002039 & 18 29 11.50 & 00 20 38.6 & 2012-06-03 & 4x600 & 4x600 & 4x600 & 089.C-0840 (Manara) \\ 
Ser34 & SSTc2d J182944.1+003356 & 18 29 44.11 & 00 33 56.0 & 2012-06-03 & 4x600 & 4x600 & 4x600 & 089.C-0840 (Manara)  \\ 
RX J1842.9-3532 & ... & 18 42 57.95 & -35 32 42.7 & 2012-05-21 & 4x300 & 4x300 & 4x300 & 089.C-0840 (Manara)  \\ 
RX J1852.3-3700 & ... & 18 52 17.29 & -37 00 11.9 & 2012-06-03 & 4x300 & 4x300 & 4x300 & 089.C-0840 (Manara) \\ 

\hline
\end{tabular}
\end{table}


\begin{table}[!]
\caption{\label{tab::Cl3}Class~III YSOs properties and observing log}
\begin{tabular}{ll | cc | c | ccc | ccccc | c }
\hline\hline

Name & Other names & RA (2000) & DEC (2000) & OBS. DATE & \multicolumn{3}{c}{T$_{\rm exp}$ (sec)} & SpT & T$_{\rm eff}$ & $A_V$ & $d$ & \lstar & Pr.Id (PI)\\
\hbox{} & \hbox{} & h m s & \degree \ \arcmin \ \arcsec & YY-MM-DD & UVB & VIS & NIR & \hbox{} & [K] & [mag] & [pc] & [\lsun] &  \hbox{}  \\

\hline

IC348-127 & Cl* IC 348 CPS 127 & 03 45 07.9 & 32 04 01.8 & 2012-11-11 & 4x150 & 4x150 & 4x150 & G4 & 5800 & 6.0 & 320 & 12.9$\pm$5.9 & 090.C-0050 (Manara)   \\ 
T21 & Ass Cha T 2-21 & 11 06 15.4 & -77 21 56.9 & 2010-01-19 & 2x70 & 2x60 & 2x60 & G5 & 5770 & 3.2 & 160 & 18.5$\pm$8.5 & 084.C-1095 (Herczeg)   \\ 
CrA75 & RX J190222.0-365541 & 19 02 22.1 & -36 55 40.9 & 2012-05-17 & 2x300 & 2x300 & 2x300 & K2 & 4900 & 1.5 & 130 & 0.4$\pm$0.2 & 089.C-0840 (Manara)   \\
\hline
\end{tabular}
\end{table}

\end{landscape}


\section{Accretion and photospheric parameters}
\label{sect::method}

In the following, we briefly describe the procedure adopted to derive self-consistently from the complete X-Shooter spectrum SpT, $A_V$, and the accretion luminosity (\lacc) for our targets. The method is described in detail in \citet{Manara13b} and is based on the fit of various parts of the observed spectra to derive these parameters. In particular, the analysis of the UV-excess together with that of absorption features at longer wavelengths allows us to determine properly the stellar properties and, at the same time, leads to an accurate and direct determination of the accretion properties. We then report the results obtained and compare these to the values derived in the literature.

\subsection{Method description}

Our method is based on a fitting procedure that considers the following three components to reproduce the observed spectrum. We include a range of photospheric template spectra - Class III YSOs from \citet{Manara13a}, augmented with some earlier SpT templates, as explained in Sect.~\ref{sec::cl3} -. We consider a range of possible values for $A_V$, and we model the excess spectrum produced by the disk accretion process with a set of isothermal hydrogen slab emission spectra. The photospheric template spectrum and the slab model are normalized using two different normalization constants determined in the fitting procedure. 

The best fit model is derived by minimizing a $\chi^2_{\rm like}$ function defined as the sum of the squared deviations (data - model) divided by the error. The $\chi^2_{\rm like}$ is computed in different regions of the UVB and VIS arms of the spectra, including the Balmer and Paschen continua region and some spectral regions around $\lambda\sim$ 700 nm characterized by molecular features particularly strong in late-type stars. We also perform a visual check of the best fit in different photospheric features sensitive to the SpT of the target and veiled by the accretion emission. 

The SpT of the best fit photospheric template is assumed for the input target with a typical uncertainty of one spectral sub-class. The best fit determined $A_V$ has an uncertainty $\lesssim$0.4 mag, which takes into account both the uncertainty on the template $A_V$ \citep[0.3 mag;][]{Manara13a} and the one on the best-fit estimate \citep[$\sim$ 0.2 mag;][]{Manara13b}. We then derive \lacc \ by integrating the normalized best fit slab model spectrum from 50 nm to 2478 nm to include all the emission of the model. This value has an estimated uncertainty of $\sim$0.2 dex \citep{Manara13b}. The value of \lstar \ is obtained from the luminosity of the best fit photospheric template after properly taking into account the normalization factor. The uncertainty on \lstar, obtained considering the one on the \lstar \ of the template \citep[$\sim$ 0.2 dex;][]{Manara13a}, on the best-fit, and on the distance, is $\sim$0.25 dex.

From the SpT of the photospheric template we derive the $T_{\rm eff}$ of the object using the SpT-$T_{\rm eff}$ relation from \citet{Luhman03} for M-type stars and \citet{KH95} for earlier SpT objects. The stellar mass ($M_*$) is then derived by interpolating evolutionary models of \citet{Baraffe98} in the position of the object on the HR Diagram, and its uncertainty is computed perturbing the position on the HR Diagram with the aforementioned uncertainty. Finally, \macc \ is derived using the classical relation \macc =  1.25 $\cdot$ \lacc $R_* /(G M_* )$ \citep{Hartmann98}, and this value has a typical uncertainty of $\sim$ 0.4 dex, obtained propagating the uncertainties on $R_*$, $M_*$, and \lacc.

We adopt the reddening curve from \citet{Cardelli}. The value of $R_V$, which is usually uncertain in young star forming regions, is assumed to be $R_V$=5.5 for CrA \citep[][and reference therein]{Peterson11} and $R_V$=3.1 for the other regions. In particular, the analysis of the best fit results for the objects located in the $\rho$-Ophiucus region show that the use of the standard extinction curve at optical wavelength is more appropriate to reproduce the observations of our targets. Also for objects located in the Chameleon~I region we find that the standard value $R_V$=3.1 better describes the observed spectrum of T21, as we reported in Sect.~\ref{sec::cl3}.

In addition to the fitting of the UVB and VIS arms of our spectra described above, we also perform a check of the best fit results using the NIR arm spectra, i.e. at $\lambda \gtrsim$ 1000 nm. For definition, TDs have low or negligible emission in excess to the photosphere at near-infrared wavelengths, and strong excess at mid-infrared and far-infrared wavelengths \citep[e.g.,][]{Calvet05}. For this reason, we expect our best fit photospheric template to match the target spectrum also in the NIR arm. This does not apply when fitting PTDs, where the contribution of inner-disk emission at near-infrared wavelengths, which is not included in our models, is not negligible. In the latter case we expect the photospheric template spectrum to lay below the target one in the near-infrared. In this check we include also, when available, the 3.6 $\mu$m Spitzer magnitude of the object, after correcting it for extinction following the prescription of \citet{McClure09}, and the magnitude of the template. The analysis of the IR color excess will be described in detail in Sect.~\ref{sec::IR_excess}. The best fit stellar and accretion parameters for the targets are reported in Table~\ref{tab::XS}.

Finally, we use the relations between the luminosity of some emission lines ($L_{\rm line}$) and $L_{\rm acc}$ calibrated by \citet{Alcala14} to verify our derived parameters. If the best fit $L_{\rm acc}$ and $A_V$ are correct, we expect to derive compatible values of $L_{\rm acc}$ from the luminosity of emission lines located in different parts of the spectra with no particular wavelength dependence. We select for this check the following 5 emission lines spread along the whole spectrum: H$\alpha$ ($\lambda$ 656.3 nm), H$\beta$ ($\lambda$ 486.1 nm), H$\gamma$ ($\lambda$ 434.0 nm), Pa$\beta$ ($\lambda$ 1281.8 nm), and Br$\gamma$ ($\lambda$ 2166.1 nm). We report the fluxes of these lines in Table~\ref{tab::lines}. We collect in the same table also the equivalent width (EW) of the lithium line at $\lambda$ 670.8 nm, that is an indicator of young ages and confirms the YSO status of all our objects.

In \citet{Manara13b}, \citet{Alcala14}, and \citet{Rigliaco12} the procedure described above has been tested on low-mass stars with SpT later than $\sim$K5. We show here its validity also for YSOs with early-K SpT. As reported in Sect.~\ref{sec::cl3}, we stress that the sample of Class~III YSOs available is highly incomplete when considering objects with SpT in the interval from G5 to K5, given that we have at disposal only one Class~III YSO with SpT K2. Finally, a more detailed analysis is needed for objects with G-type SpT, the so-called intermediate mass stars, because for these objects the excess emission due to accretion can be hardly detected in the wavelength region covered by X-Shooter \citep[e.g.][]{Calvet04}. We thus discuss in the following sections these three different types of objects separately.

\subsubsection{Results for low mass stars} 
\begin{figure*}
\includegraphics[width=0.9\textwidth]{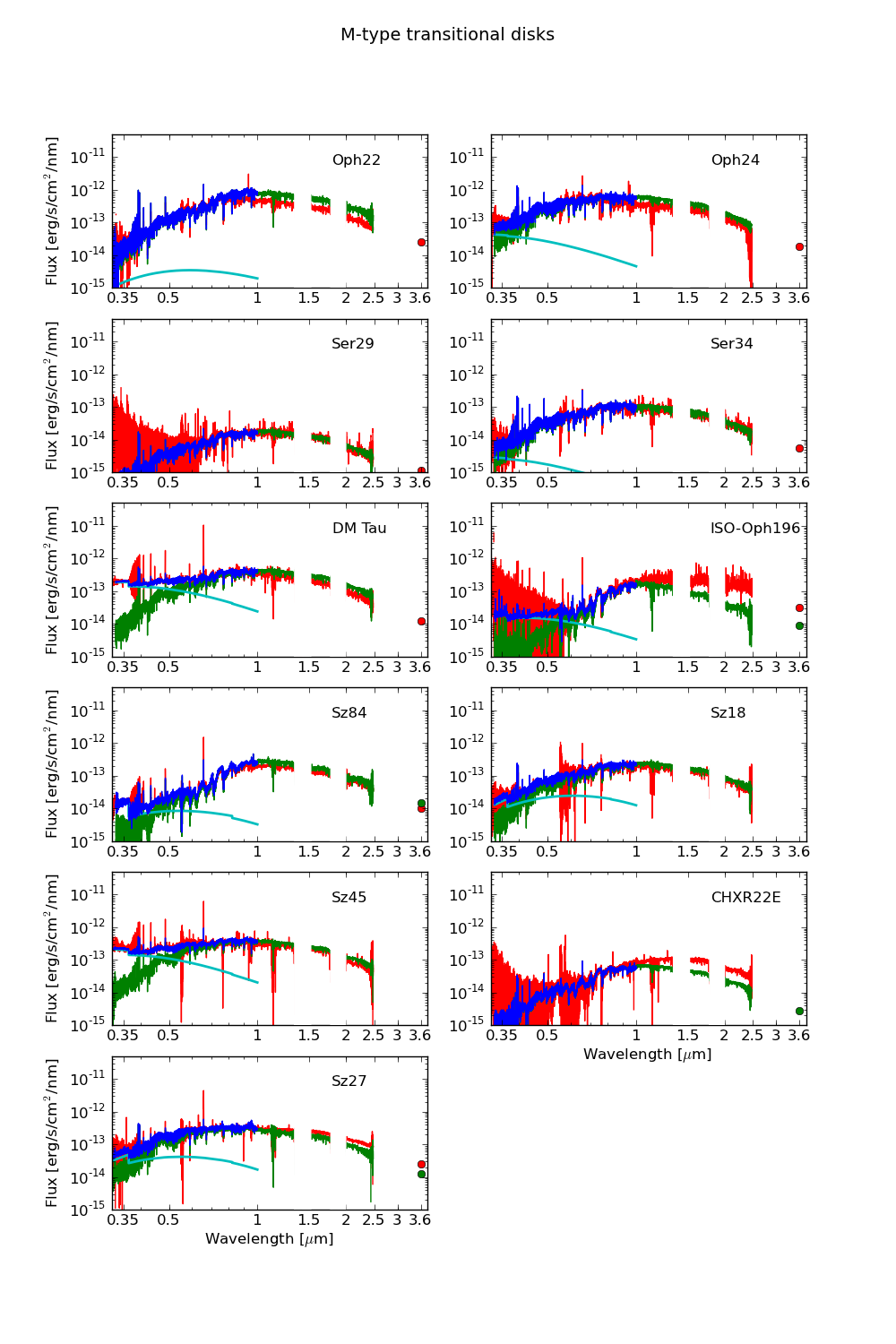}
\caption{Best fit for transitional disks with spectral type in the M class. The red line is the observed dereddened spectrum, green line the photospheric template, light blue line the slab model, and blue line the best fit. }
        \label{fig::VLM_acc}
\end{figure*}

\begin{figure*}
\includegraphics[width=\textwidth]{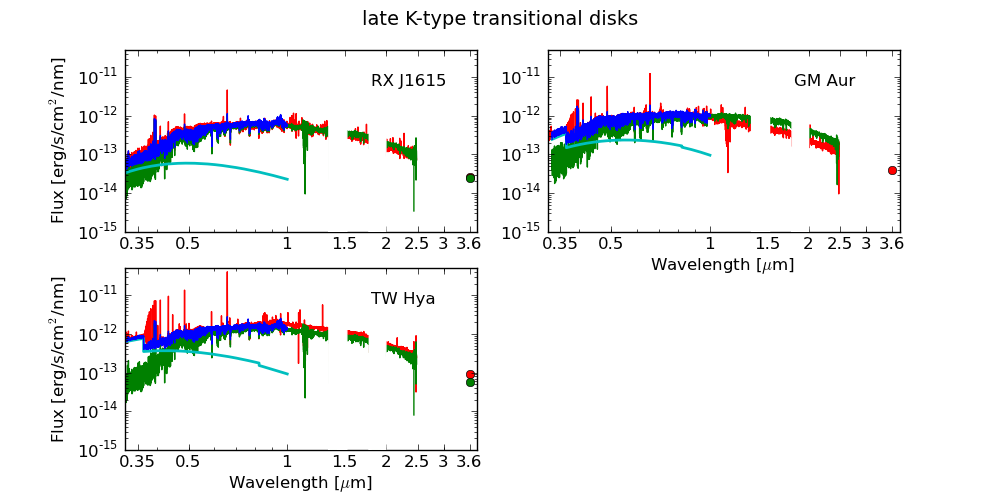}
\caption{Best fit for late-K type transitional disks. Colors as in Fig.~\ref{fig::VLM_acc} }
        \label{fig::lateK_acc}
\end{figure*}

The best fits obtained for TDs with M SpT are shown in Fig.~\ref{fig::VLM_acc}, while those for TDs with SpT later or equal to K5 in Fig.~\ref{fig::lateK_acc}. The observed and reddening corrected spectra are shown with a red line, the green line represents the photospheric templates used, the light blue line the slab model, and the blue line the best fit, which is the sum of the photospheric template and the slab model. The best fit is plotted only in the regions where it is calculated, i.e. $\lambda\lambda \sim$330-1000 nm. The agreement between the best fit and the observed spectrum in this wavelength range is always very good. At wavelengths longer than $\sim$ 1000 nm we plot only the photospheric template and the observed spectra, including also their 3.6 $\mu$m Spitzer magnitudes, when available. As mentioned earlier, we expect the photospheric template spectrum not to exceed the target one in this region. This is the case for most of the targets, but not for Oph22, Oph24, DM Tau, and GM Aur. For one object, Sz27, the excess emission at near-infrared wavelengths confirms the previous classification as PTD. According to our best fits, also CHXR22E and ISO-Oph 196 should be classified as PTD. 

The stellar and accretion parameters obtained for these targets are reported in Table~\ref{tab::XS}. For the objects considered in this section, i.e. with SpT later or equal to K5, the best fit SpT is the same within up to one or two spectral sub-class as the one reported in the literature. In most cases the difference with respect to the literature values is small also for the other stellar and accretion parameters. In particular, values of \macc \ agree within 0.3 dex with those reported in the literature. The objects with larger differences are Sz18, Sz45, RX J1615, Oph24, Ser29, and Ser34. We suggest that these differences are due to the different methodologies of previous studies with respect to our. Variability of accretion would result in a smaller difference. Recent studies showed that in most young accreting stars these variations are in general smaller than 0.3 dex \citep[e.g.,][]{Costigan12}. For Ser29 we are only able to provide an upper limit on \lacc \ from the fitting due to the low signal-to-noise of the spectrum in the whole UVB arm. This value is compatible with the measurement of the H$\alpha$ line, which is the only line seen in emission in the spectrum.

\begin{table*} 
\centering 
\caption{\label{tab::XS}Stellar, disk, and accretion parameters of the targets} 
\begin{tabular}{lccccccccccccc} 
\hline\hline 
Name & dist & R$_{\rm in}$ & SpT   & T$_{\rm eff}$ & A$_V$ & L$_*$ & log \lacc & M$_*$ & log \macc & $R_*$  & Disk & \multicolumn{2}{c}{Ref}  \\ 
\hbox{} & [pc] & [AU] & \hbox{}  &  [K]  & [mag] & [L$_\odot$] & [L$_\odot$] & [M$_\odot$] & [M$_\odot$/yr] & [$R_\odot$] & type  & \hbox{} & \hbox{}\\ 
\hline 
LkH$\alpha$330 & 250 & 68 & G4 & 5800 & 3.0 & 14.40 & -0.6 & 2.35$\pm$0.57 & -7.8 & 3.74$\pm$1.08 & PTD & 1 & D \\ 
DM Tau & 140 & 19 & M2 & 3560 & 1.1 & 0.36 & -1.3 & 0.56$\pm$0.08 & -8.2 & 1.57$\pm$0.46 & TD & 1 & B \\ 
LkCa 15 & 140 & 50 & K2 & 4900 & 1.2 & 1.21 & -1.1 & 1.24$\pm$0.33 & -8.4 & 1.52$\pm$0.44 & PTD & 1 & B \\ 
GM Aur & 140 & 28 & K5 & 4350 & 0.6 & 0.99 & -1.0 & 1.36$\pm$0.36 & -8.3 & 1.75$\pm$0.51 & PTD & 1 & B \\ 
Sz-Cha & 160 & 29 & K2 & 4900 & 1.3 & 1.17 & -0.5 & 1.22$\pm$0.32 & -7.8 & 1.50$\pm$0.43 & PTD & 2 & B \\ 
TW Hya & 55 & 4 & K7 & 4060 & 0.0 & 0.18 & -1.6 & 0.79$\pm$0.17 & -8.9 & 0.85$\pm$0.25 & TD & 3 & B \\ 
CS Cha & 160 & 43 & K2 & 4900 & 0.8 & 1.45 & -1.0 & 1.32$\pm$0.37 & -8.3 & 1.66$\pm$0.48 & TD & 4 & B \\ 
CHXR22E & 160 & 7 & M4 & 3270 & 2.6 & 0.07 & -4.1 & 0.24$\pm$0.06 & -10.9 & 0.82$\pm$0.24 & PTD & 2 & B \\ 
Sz18 & 160 & 8 & M2 & 3560 & 1.3 & 0.26 & -1.9 & 0.54$\pm$0.08 & -8.9 & 1.34$\pm$0.39 & TD & 2 & B \\ 
Sz27 & 160 & 15 & K7 & 4060 & 2.9 & 0.33 & -1.6 & 0.96$\pm$0.24 & -8.9 & 1.16$\pm$0.34 & PTD & 2 & B \\ 
Sz45 & 160 & 18 & M0.5 & 3780 & 0.7 & 0.42 & -1.2 & 0.85$\pm$0.11 & -8.3 & 1.51$\pm$0.44 & TD & 2 & B \\ 
Sz84 & 150 & 55 & M5 & 3125 & 0.5 & 0.24 & -2.3 & 0.24$\pm$0.06 & -8.9 & 1.67$\pm$0.49 & TD & 5 & B \\ 
RX J1615 & 185 & 30 & K7 & 4060 & 0.0 & 0.89 & -1.3 & 1.16$\pm$0.16 & -8.5 & 1.90$\pm$0.55 & TD & 1 & B \\ 
Oph22 & 125 & 1 & M3 & 3415 & 3.0 & 0.56 & -2.9 & 0.53$\pm$0.14 & -9.7 & 2.13$\pm$0.62 & TD & 5 & B \\ 
Oph24 & 125 & 3 & M0 & 3850 & 4.0 & 0.42 & -2.0 & 0.92$\pm$0.13 & -9.2 & 1.45$\pm$0.42 & TD & 5 & B \\ 
SR 21 & 125 & 36 & G4 & 5800 & 6.0 & 8.11 & -0.7$^a$ & 1.95$\pm$0.50 & -7.9$^a$ & 2.81$\pm$0.81 & PTD & 1 & D \\ 
ISO-Oph196 & 125 & 15 & M5.5 & 3060 & 3.0 & 0.08 & -2.3 & 0.14$\pm$0.04 & -8.9 & 1.00$\pm$0.29 & PTD & 1 & B \\ 
DoAr 44 & 125 & 30 & K2 & 4900 & 1.7 & 0.64 & -0.9 & 0.97$\pm$0.19 & -8.2 & 1.11$\pm$0.32 & PTD & 1 & B \\ 
Ser29 & 230 & 8 & M2 & 3560 & 2.6 & 0.04 & $<$-3.8 & 0.47$\pm$0.08 & $<$-11.2 & 0.52$\pm$0.15 & TD & 5 & B \\ 
Ser34 & 230 & 25 & M1 & 3705 & 2.7 & 0.26 & -2.7 & 0.71$\pm$0.08 & -9.8 & 1.23$\pm$0.36 & TD & 5 & B \\ 
RX J1842.9 & 130 & 5 & K2 & 4900 & 0.4 & 0.56 & -1.5 & 0.93$\pm$0.16 & -8.8 & 1.03$\pm$0.30 & PTD & 6 & B \\ 
RX J1852.3 & 130 & 16 & K2 & 4900 & 1.0 & 0.77 & -1.4 & 1.04$\pm$0.19 & -8.7 & 1.21$\pm$0.35 & TD & 6 & B \\ 
\hline 
\end{tabular} 
\tablefoot{Reference for \rin: (1) \citet{Andrews11}, (2) \citet{Kim09}, (3) \citet{Hughes07}, (4) \citet{Espaillat13}, (5) \citet{Merin10}, (6) \citet{Hughes10},
 Evolutionary models used to derive $M_*$ and \macc: (B) \citet{Baraffe98}, (D) \citet{DAntona}. $^a$Highly uncertain value.} 
\end{table*}

\subsubsection{Results for early K-type stars} 
\begin{figure*}
\includegraphics[width=\textwidth]{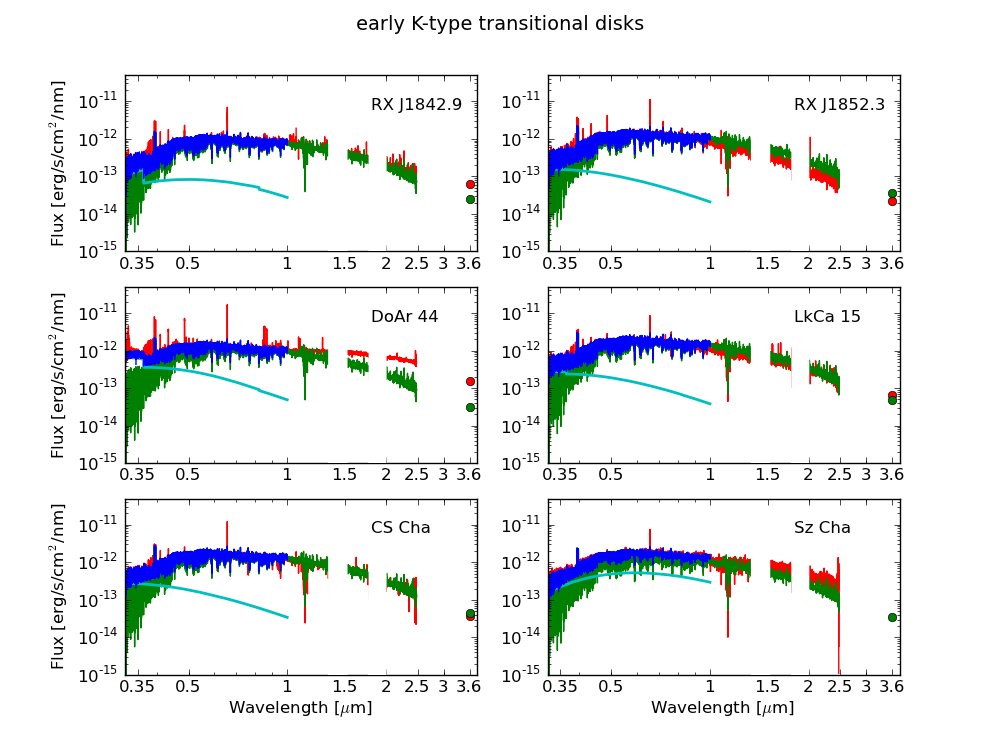}
\caption{Best fit of early K-type transitional disks. Colors as in Fig.~\ref{fig::VLM_acc} }
        \label{fig::earlyK_acc}
\end{figure*}

For six of our objects we obtain a best fit using the Class~III YSO template with SpT K2. These results are also reported in Table~\ref{tab::XS}, while their best fit are shown in Fig.~\ref{fig::earlyK_acc}. In all these cases the best fit is very good with the only exception of RX J1852.3. With respect to the literature, typical differences of the SpT from the best fit K2 are of up to two spectral sub-classes apart from CS Cha, which was previously classified as K6. We adopt for all these targets SpT K2 in our analysis, with the caveat that the uncertainty on this parameter is larger for these objects with respect to later SpT targets due to the already mentioned incompleteness of photospheric templates of SpT late-G and early-K. Our best fits confirm that DoAr44, LkCa15, and SzCha are PTD (see also Sect.~\ref{sec::IR_excess}). We find a hint of excess in the K-band spectrum of RX J1842.9, which becomes clearer at the Spitzer [3.6] data point. This confirms the observations of infrared excess in this object reported in \citet{Hughes10} and implies that also this object is a PTD. The largest difference in the derived values of \macc \ is for SzCha, which results to be a stronger accretor than previously determined.

\subsubsection{Results for intermediate-mass stars} 
\begin{figure*}
\includegraphics[width=\textwidth]{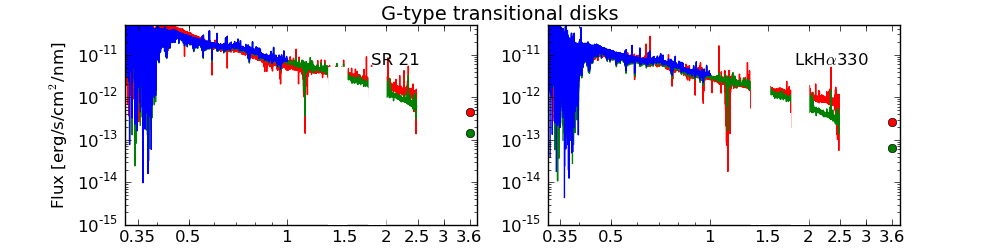}
\caption{Best fit of G-type transitional disks. Colors as in Fig.~\ref{fig::VLM_acc}.  }
        \label{fig::G_acc}
\end{figure*}

Two objects in our sample are of early-G SpT, namely LkH$\alpha$330 and SR21. For these TDs we have not been able to detect excess emission with our fitter. As also \citet{Calvet04} pointed out, the excess emission for intermediate-mass stars like these two is hard to be detected at $\lambda>$330 nm due to the similar temperatures of the accretion shock and the stellar photosphere. We have only been able to fit these spectra to derive their $A_V$ and $L_*$, and we show these best fits in Fig.~\ref{fig::G_acc}. Their positions on the HRD are not covered by the evolutionary tracks of \citet{Baraffe98}, so we derive the values of $M_*$ for these two targets using the models of \citet{DAntona}. In both objects we detect an excess emission in the near-infrared wavelengths which could imply these objects are PTD.

In the spectrum of LkH$\alpha$330 various emission lines are present, such as the H$\alpha$, H$\beta$, Pa$\beta$, and Br$\gamma$. The only \lacc-$L_{\rm line}$ relation available for this class of objects is the one reported in \citet{Calvet04} for the Br$\gamma$ line. We use this relation to derive a value of \lacc$\sim 0.23$ \lsun \ which leads to a value of \macc \ consistent with those reported in the literature. 

The hydrogen recombination lines of SR21 appear in absorption in the whole spectrum. The same is found when looking at the CaII IRT lines. Moreover, the photospheric lines of this object appear much broader than the corresponding Class~III YSO spectrum. Nevertheless, the wings of the hydrogen lines, in particular those of the H$\alpha$ line, appear in emission at very high velocities up to $\sim$ 250 km/s, thus suggesting that they originate in an accretion-related infall region. Therefore we classify this object as an accreting TD. Given that no Br$\gamma$ emission is detected in this spectrum, we derive \lacc \ from the luminosity of the H$\alpha$ line. This is derived from the dereddened spectrum corrected for the photospheric line contribution, which is estimated using a synthetic spectrum of the same $T_{\rm eff}$ and broaden to match photosferic lines close to the H$\alpha$. To convert the luminosity of the H$\alpha$ line in \lacc \ we use the relation provided by \citet{Alcala14}. Given all the assumptions adopted to estimate this value we consider the derived \lacc \ very uncertain.

\begin{table*} 
\centering 
\caption{\label{tab::lines}Derived properties of analyzed lines} 
\begin{tabular}{lcccccc} 
\hline\hline 
Name & F$_{\rm H\alpha}$ & F$_{\rm H\beta}$ & F$_{\rm H\gamma}$ & F$_{\rm Pa\beta}$ & F$_{\rm Br\gamma}$ & EW$_{\rm Li_{\lambda 670.8}}$ \\ 
\hbox{} & [erg s$^{-1}$ cm$^{-2}$] & [erg s$^{-1}$ cm$^{-2}$] & [erg s$^{-1}$ cm$^{-2}$] & [erg s$^{-1}$ cm$^{-2}$] & [erg s$^{-1}$ cm$^{-2}$] & [m\AA] \\ 
\hline 
LkH$\alpha$330 & $(1.11 \pm 0.02)  \times 10^{-11}$  & $(6.6 \pm 2.9)  \times 10^{-13}$  & $<3.2 \times 10^{-13}$  & $(6.5 \pm 0.3)  \times 10^{-13}$  & $(5.9 \pm 3.4)  \times 10^{-14}$  & 110$\pm$2  \\ 
DM Tau & $(4.48 \pm 0.02)  \times 10^{-12}$  & $(3.5 \pm 0.1)  \times 10^{-13}$  & $(2.2 \pm 0.1)  \times 10^{-13}$  & $(1.06 \pm 0.07)  \times 10^{-13}$  & $(1.4 \pm 0.2)  \times 10^{-14}$  & 410$\pm$21  \\ 
LkCa 15 & $(3.1 \pm 0.1)  \times 10^{-12}$  & $(3.7 \pm 1.6)  \times 10^{-13}$  & $(1.5 \pm 0.1)  \times 10^{-13}$  & $(1.2 \pm 0.2)  \times 10^{-13}$  & $(1.9 \pm 1.5)  \times 10^{-14}$  & 460$\pm$23  \\ 
GM Aur & $(1.06 \pm 0.02)  \times 10^{-11}$  & $(2.1 \pm 0.07)  \times 10^{-12}$  & $(7.9 \pm 0.4)  \times 10^{-13}$  & $(1.01 \pm 0.02)  \times 10^{-12}$  & $(1.6 \pm 0.1)  \times 10^{-13}$  & 440$\pm$22  \\ 
Sz Cha & $(2.6 \pm 0.1)  \times 10^{-12}$  & $<4.7 \times 10^{-14}$  & $<4.0 \times 10^{-14}$  & $(1.9 \pm 0.3)  \times 10^{-13}$  & $<1.3 \times 10^{-14}$  & 350$\pm$10  \\ 
TW Hya & $(2.39 \pm 0.04)  \times 10^{-11}$  & $(4.52 \pm 0.07)  \times 10^{-12}$  & $(2.2 \pm 0.05)  \times 10^{-12}$  & $(2.21 \pm 0.04)  \times 10^{-12}$  & $(2.61 \pm 0.08)  \times 10^{-13}$  & 430$\pm$21  \\ 
CS Cha & $(5.23 \pm 0.09)  \times 10^{-12}$  & $(4.7 \pm 1.1)  \times 10^{-13}$  & $(2.2 \pm 0.4)  \times 10^{-13}$  & $(1.1 \pm 0.1)  \times 10^{-13}$  & $(2.6 \pm 1.1)  \times 10^{-14}$  & 510$\pm$28  \\ 
CHXR22E & $(1.1 \pm 0.1)  \times 10^{-14}$  & $(4.4 \pm 0.6)  \times 10^{-15}$  & $(9.1 \pm 4.9)  \times 10^{-16}$  & $<9.1 \times 10^{-15}$  & $<2.4 \times 10^{-15}$  & 230$\pm$56  \\ 
Sz18 & $(3.8 \pm 0.2)  \times 10^{-13}$  & $(2.5 \pm 0.6)  \times 10^{-14}$  & $(7.9 \pm 0.9)  \times 10^{-15}$  & $(4.3 \pm 1.7)  \times 10^{-15}$  & $<2.7 \times 10^{-15}$  & 590$\pm$77  \\ 
Sz27 & $(1.83 \pm 0.08)  \times 10^{-12}$  & $(7.8 \pm 0.6)  \times 10^{-14}$  & $(3.2 \pm 0.2)  \times 10^{-14}$  & $(4.9 \pm 0.5)  \times 10^{-14}$  & $(1.0 \pm 0.2)  \times 10^{-14}$  & 510$\pm$29  \\ 
Sz45 & $(2.83 \pm 0.02)  \times 10^{-12}$  & $(4.6 \pm 0.1)  \times 10^{-13}$  & $(2.9 \pm 0.1)  \times 10^{-13}$  & $(1.17 \pm 0.08)  \times 10^{-13}$  & $(2.0 \pm 0.5)  \times 10^{-14}$  & 440$\pm$32  \\ 
Sz84 & $(7.0 \pm 0.1)  \times 10^{-13}$  & $(5.59 \pm 0.09)  \times 10^{-14}$  & $(2.7 \pm 0.04)  \times 10^{-14}$  & $(3.1 \pm 0.3)  \times 10^{-14}$  & $(1.1 \pm 0.5)  \times 10^{-14}$  & 460$\pm$20  \\ 
RX J1615 & $(2.24 \pm 0.05)  \times 10^{-12}$  & $(2.5 \pm 0.3)  \times 10^{-13}$  & $(6.9 \pm 1.1)  \times 10^{-14}$  & $(7.5 \pm 1.6)  \times 10^{-14}$  & $(1.5 \pm 0.7)  \times 10^{-14}$  & 550$\pm$32  \\ 
Oph22 & $(1.9 \pm 0.09)  \times 10^{-13}$  & $(5.4 \pm 0.1)  \times 10^{-14}$  & $(2.8 \pm 0.1)  \times 10^{-14}$  & $<1.7 \times 10^{-14}$  & $<7.3 \times 10^{-15}$  & 570$\pm$40  \\ 
Oph24 & $(3.2 \pm 0.2)  \times 10^{-13}$  & $(1.08 \pm 0.03)  \times 10^{-13}$  & $(7.7 \pm 0.3)  \times 10^{-14}$  & $<6.4 \times 10^{-14}$  & $(5.0 \pm 1.5)  \times 10^{-15}$  & 570$\pm$37  \\ 
SR 21$^a$ & ...  & ...  & ...  & ...  & ...  & 140$\pm$2  \\ 
ISO$-$Oph196 & $(3.54 \pm 0.01)  \times 10^{-13}$  & $(8.8 \pm 0.6)  \times 10^{-14}$  & $(8.8 \pm 0.8)  \times 10^{-14}$  & $<7.4 \times 10^{-14}$  & $(2.2 \pm 0.3)  \times 10^{-14}$  & 370$\pm$31  \\ 
DoAr 44 & $(9.4 \pm 0.1)  \times 10^{-12}$  & $(2.66 \pm 0.08)  \times 10^{-12}$  & $(1.04 \pm 0.06)  \times 10^{-12}$  & $(1.12 \pm 0.01)  \times 10^{-12}$  & $(2.6 \pm 0.1)  \times 10^{-13}$  & 420$\pm$19  \\ 
Ser29 & $(1.3 \pm 0.1)  \times 10^{-14}$  & $<1.7 \times 10^{-15}$  & $<8.0 \times 10^{-15}$  & $<1.9 \times 10^{-15}$  & $<4.8 \times 10^{-17}$  & 380$\pm$199  \\ 
Ser34 & $(1.02 \pm 0.05)  \times 10^{-13}$  & $(7.8 \pm 0.7)  \times 10^{-15}$  & $(5.9 \pm 1.1)  \times 10^{-15}$  & $<1.6 \times 10^{-15}$  & $(1.8 \pm 0.4)  \times 10^{-15}$  & 630$\pm$40  \\ 
RX J1842.9 & $(3.41 \pm 0.08)  \times 10^{-12}$  & $(4.6 \pm 0.6)  \times 10^{-13}$  & $(2.0 \pm 0.2)  \times 10^{-13}$  & $(1.2 \pm 0.1)  \times 10^{-13}$  & $(1.5 \pm 0.7)  \times 10^{-14}$  & 440$\pm$21  \\ 
RX J1852.3 & $(5.67 \pm 0.09)  \times 10^{-12}$  & $(1.01 \pm 0.07)  \times 10^{-12}$  & $(2.8 \pm 0.4)  \times 10^{-13}$  & $(2.0 \pm 0.2)  \times 10^{-13}$  & $(4.0 \pm 2.5)  \times 10^{-14}$  & 510$\pm$30  \\ 
\hline 
\end{tabular} 
\tablefoot{Fluxes are reported in the format (flux $\pm$ err) multiplied by the order of magnitude. $^a$The estimate of the flux of the emission lines of SR21 is very uncertain, thus we do not report these values for this object.} 
\end{table*}

\subsection{Infrared color excess}
\label{sec::IR_excess}

\begin{figure}
\includegraphics[width=0.5\textwidth]{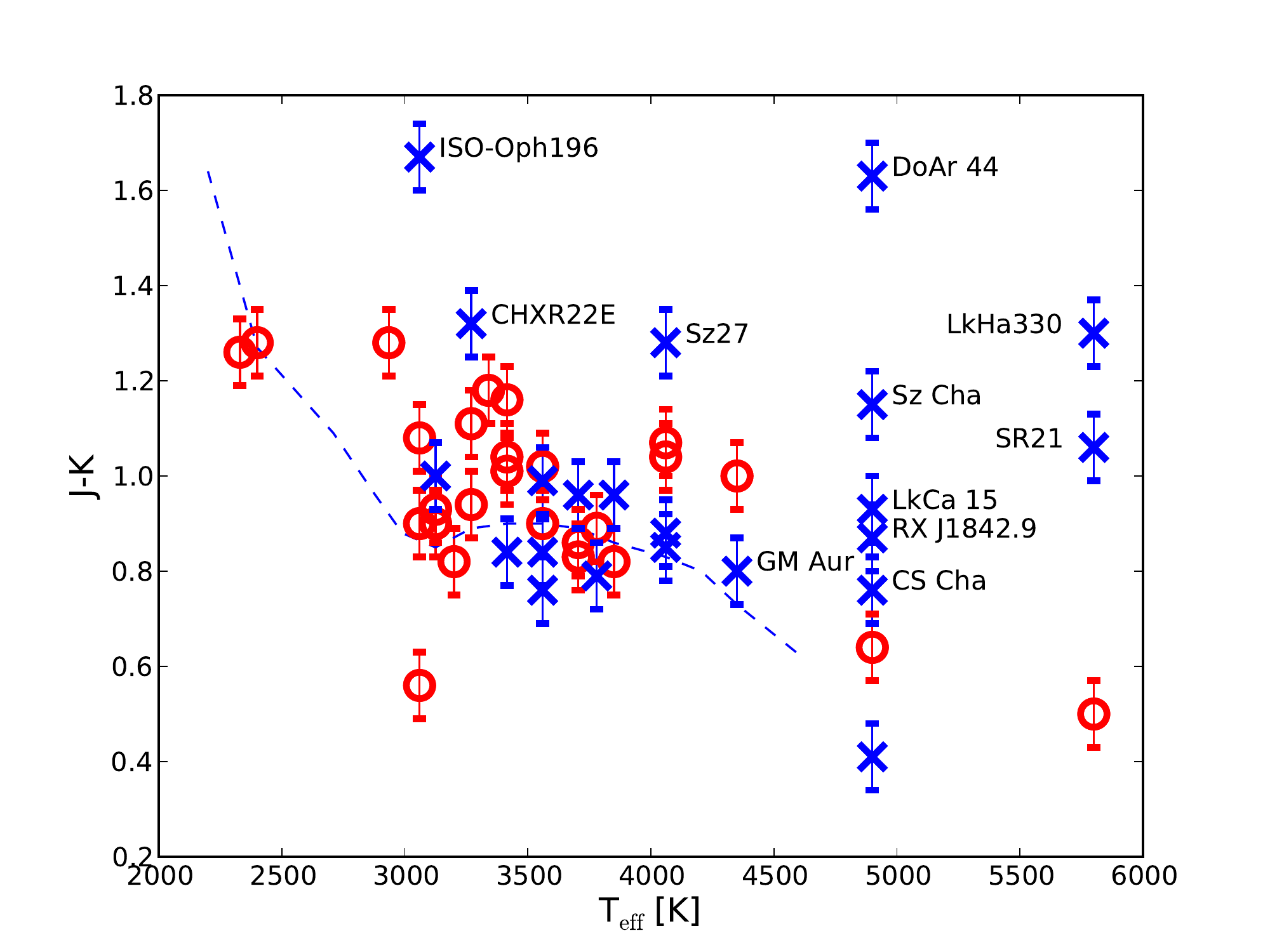}
\caption{$J-K$ color calculated with synthetic photometry on the best fit dereddened TD spectra ({\it blue crosses}) vs $T_{\rm eff}$ of the targets. The {\it red circles} represent the Class~III YSOs colors derived with synthetic photometry on their spectra. The dashed line represents the photospheric color of YSOs according to \citet{Luhman10}.}
        \label{fig::IR_color1}
\end{figure}

\begin{figure}
\includegraphics[width=0.5\textwidth]{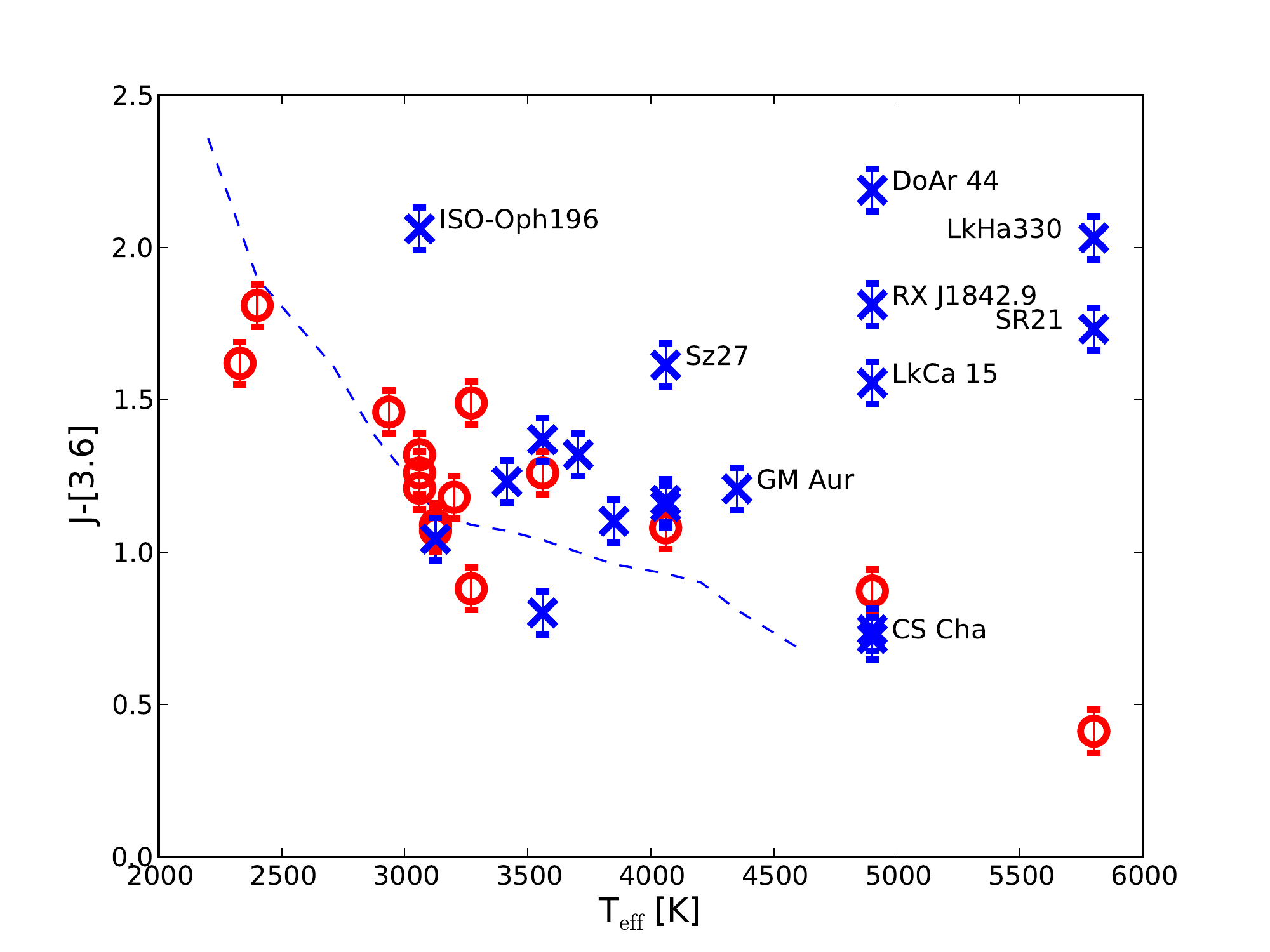}
\caption{$J-$[3.6] color vs $T_{\rm eff}$ of the targets. The $J$ magnitude is calculated on the best-fit dereddened TD spectra ({\it blue crosses}), while the [3.6] magnitude is derived from the literature. Colors and symbols are the same as in Fig.~\ref{fig::IR_color1}.}
        \label{fig::IR_color2}
\end{figure}

From the best fit derived as explained above we analyze the color excess in the IR colors in order to detect emission from the innermost dusty disk. We perform synthetic photometry on the dereddened TD spectra and on the Class~III YSO spectra. We then plot in Fig.~\ref{fig::IR_color1}-\ref{fig::IR_color2} the $J-K$ and $J-$[3.6] colors as a function of $T_{\rm eff}$ both for the Class~III YSOs ({\it red circles}) and the TDs ({\it blue crosses}). These colors trace the presence of an inner disk, which would result in an excess emission with respect to the photosphere in the $K$ band and at 3.6 $\mu$m. As a reference, we plot with a dashed line the photospheric color locus of diskless YSOs derived by \citet{Luhman10}. We see that the colors of the Class~III YSOs are distributed on these plots around this empirically calibrated locus with a small dispersion. Also most of the TDs have colors compatible with the Class~III YSOs at the same $T_{\rm eff}$, meaning that their IR colors are compatible with photospheric ones, thus they have no dust-rich inner disk. In some cases, however, the excess is detectable and the objects should be classified as PTD. This is the case for the objects already listed in the previous sections, namely ISO Oph 196, CHXR22E, Sz 27, DoAr 44, Sz Cha, LkCa15, RX J1842.9, LkH$\alpha$330, and SR21, and for GM Aur, given the excess in the $J-$[3.6] color (see Fig.~\ref{fig::IR_color2}). This object was also previously classified as PTD by \citet{Calvet05} and \citet{Espaillat10}. For one object, CS Cha, the excess is detected in the $J-K$ color but not in the $J$-[3.6] one, and in the former it is compatible with the Class~III YSOs color. We thus classify this object as TD.  We report this classification in Table~\ref{tab::XS}. The objects classified as PTD have $R_{\rm in}$ values that range smoothly from 5 to 68 AU. The presence of a dusty innermost region of the disk is thus uncorrelated with the size of the dust-depleted gap.

\section{Wind signatures}
\label{sect::wind}

The most prominent forbidden line present in the spectra of our TDs is the [OI] $\lambda$ 630 nm line. This line has been detected in many accreting YSOs \citep[e.g.,][]{Hartigan95} and can present two distinct components. The high-velocity component (HVC, $\Delta v\sim$ 100-200 km s$^{-1}$) of this line is known to trace collimated jets. The origin of the low-velocity component (LVC, $\Delta v\sim$ 2-3 km s$^{-1}$), instead, is still unclear. It is believed to originate in the disk or from the base of a slow disk wind \citep{Hartigan95}, but there are suggestions that it can be originated in a photoevaporative wind \citep[][and references therein]{Rigliaco13}. We detect the LVC of this line in 17 ($\sim$80 \%) of the spectra of our TDs, with a clear detection of the HVC only in ISO-Oph 196. 

\begin{figure*}[h!t]
\includegraphics[width=\textwidth]{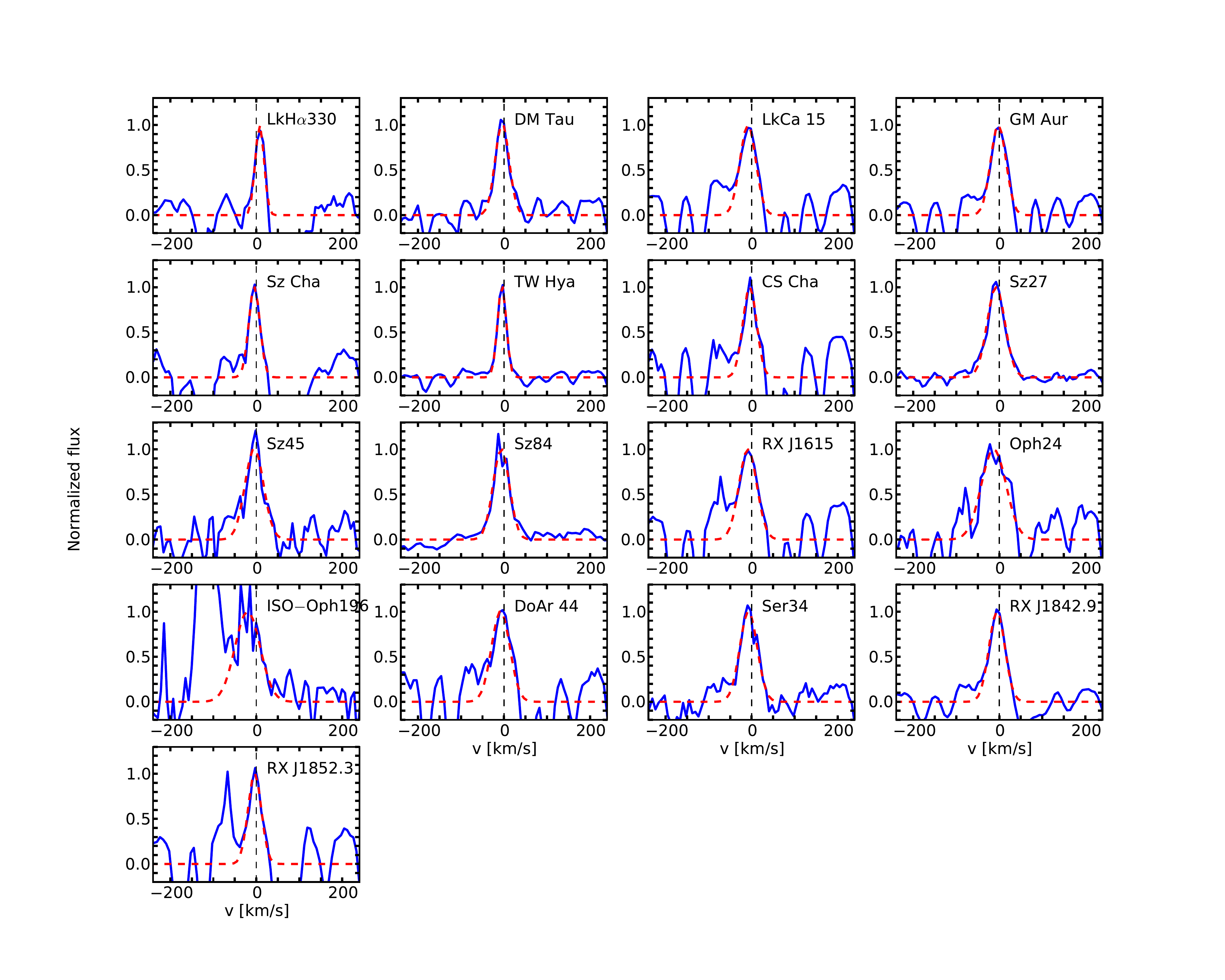}
\caption{Normalized [OI] 630 nm line for the TDs in our sample where this line has been detected. The red dashed line is the best gaussian fit of the low-velocity component of the line. }
        \label{fig::OI}
\end{figure*}

We derive the flux and the peak velocity of the LVC of the [OI]$\lambda$ 630 nm line in the following way. We firstly refine the wavelength calibration by fitting the photospheric Li~I line at $\lambda$ 670.78 nm and shifting the spectra to match the nominal central wavelength of this line. This line is detected in all the objects and its EW are reported in Column~7 of Table~\ref{tab::lines}. Then, we fit with a gaussian profile on the dereddend spectrum the LVC of the [OI]$\lambda$630 nm line and we integrate the flux of the best fit to derive the line flux. The error on the flux is derived from the standard deviation of the continuum estimated around the line. The derived flux, error, peak velocity ($v_0$), and FWHM of the gaussian fit is reported in Table~\ref{tab::forb_lines}. The lines and their best fits are shown in Fig.~\ref{fig::OI}. In all the objects with detected [OI]$\lambda$630 nm line, with the exception of LkH$\alpha$330, the line is slightly blueshifted, with values of $v_0$ ranging from $\sim$ -2 km/s to $\sim$ -8 km/s in most cases, and only two objects (Oph24 and ISO-Oph 196) with $v_0 <$ -10 km/s. Even if the exact value of $v_0$ in each object is still uncertain also after the procedure to correct the wavelength calibration described above, we see that the [OI]$\lambda$630 nm line is systematically blueshifted, meaning that it is originated in some kind of wind. The mean value of the FWHM of the [OI]$\lambda$630 nm line derived from the spectra of our targets is $\sim$ 40 km/s. We note, however, that the values of FWHM $\lesssim$ 30 km/s should be considered with caution, as these values are close to the nominal resolution of the instrument.

\begin{table} 
\centering 
\caption{\label{tab::forb_lines}Derived properties of [OI] line at $\lambda$ 630 nm} 
\begin{tabular}{lccc} 
\hline\hline 
Name & F$_{\rm [OI]\lambda 630}$ & $v_{0,\rm [OI]\lambda 630}$ & FWHM$_{\rm [OI]\lambda 630}$ \\ 
\hbox{} & [erg s$^{-1}$ cm$^{-2}$] & [km/s] & [km/s] \\ 
\hline 
LkH$\alpha$330 & $(6.1 \pm 1.9)  \times 10^{-14}$  & 7.7 & 24  \\ 
DM Tau & $(9.1 \pm 1.8)  \times 10^{-15}$  & -4.3 & 38  \\ 
LkCa 15 & $(4.6 \pm 1.9)  \times 10^{-14}$  & -8.7 & 43  \\ 
GM Aur & $(3.9 \pm 1.0)  \times 10^{-14}$  & -2.2 & 42  \\ 
Sz Cha & $(2.4 \pm 0.9)  \times 10^{-14}$  & -4.7 & 30  \\ 
TW Hya & $(9.0 \pm 0.8)  \times 10^{-14}$  & -4.8 & 24  \\ 
CS Cha & $(2.8 \pm 1.5)  \times 10^{-14}$  & -4.4 & 37  \\ 
CHXR22E & $<8.0 \times 10^{-16}$  & ... & ...  \\ 
Sz18 & $<2.6 \times 10^{-15}$  & ... & ...  \\ 
Sz27 & $(5.4 \pm 0.3)  \times 10^{-14}$  & -7.6 & 49  \\ 
Sz45 & $(1.8 \pm 0.5)  \times 10^{-14}$  & -4.8 & 50  \\ 
Sz84 & $(2.4 \pm 0.2)  \times 10^{-15}$  & -6.5 & 44  \\ 
RX J1615 & $(1.7 \pm 0.8)  \times 10^{-14}$  & -8.7 & 49  \\ 
Oph22 & $<7.3 \times 10^{-15}$  & ... & ...  \\ 
Oph24 & $(3.0 \pm 1.3)  \times 10^{-14}$  & -13.0 & 67  \\ 
SR 21 & $<6.4 \times 10^{-14}$  & ... & ...  \\ 
ISO$-$Oph196 & $(1.4 \pm 0.4)  \times 10^{-15}$  & -20.4 & 68  \\ 
DoAr 44 & $(2.6 \pm 1.2)  \times 10^{-14}$  & -7.7 & 50  \\ 
Ser29 & $<4.2 \times 10^{-16}$  & ... & ...  \\ 
Ser34 & $(4.1 \pm 0.8)  \times 10^{-15}$  & -7.3 & 47  \\ 
RX J1842.9 & $(5.0 \pm 0.8)  \times 10^{-14}$  & -5.5 & 43  \\ 
RX J1852.3 & $(2.6 \pm 1.3)  \times 10^{-14}$  & -3.8 & 36  \\ 
\hline 
\end{tabular} 
\tablefoot{Fluxes are reported in the format (flux $\pm$ err) multiplied by the order of magnitude.}
\end{table}

\section{Discussion}
\label{sect::disc}

In this section we discuss the accretion and wind properties of our targets and we estimate the amount of gaseous material in their inner disks. It should be kept in mind that our sample is composed mostly by objects already known to be strong accretors and it is {\it not} an unbiased sample. Nevertheless, the properties of these strong-accreting TDs have important consequences on our understanding of the TDs formation and evolution, as we discuss in the following.

\subsection{Accretion properties}\label{sect::accretion}

Here we aim at understanding whether there is a dependence of the accretion properties of our objects with the morphology of the disk, in particular with \rin, and whether there are differences with respect to accretion properties in cTTs.

\begin{figure}
\centering
\includegraphics[width=0.5\textwidth]{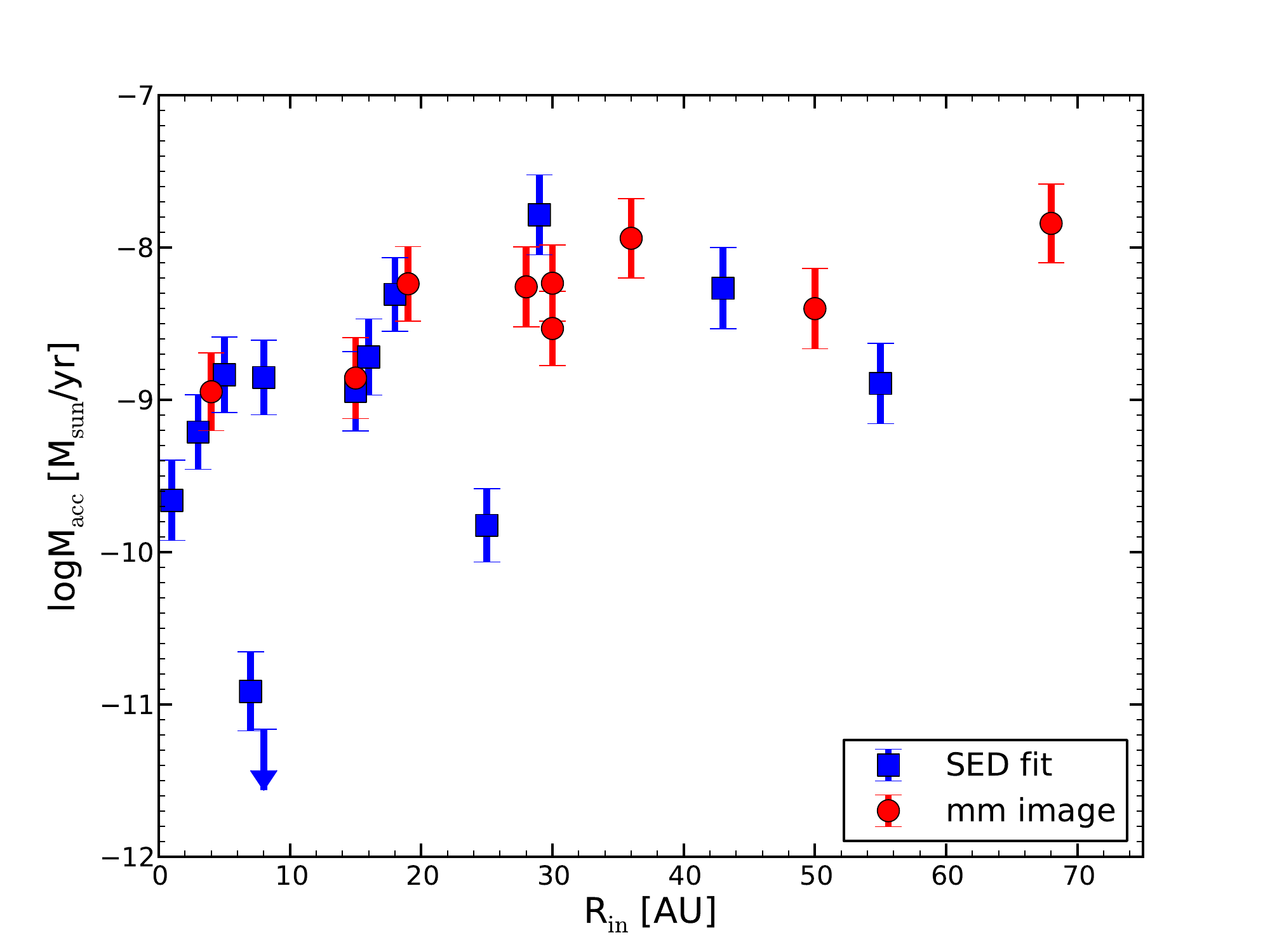}
\caption{Logarithm of the mass accretion rate vs inner hole size for our sample. Different symbols are used to distinguish the methods used in the literature to derive the size of the inner hole. {\it Blue squares} are adopted when this has been derived using IR-SED fitting, while {\it red circles} when the values are derived from resolved mm-interferometry observations. Downward arrows are upper limits. The two lowest points are, from left to right, CHXR22E and Ser29. The object at \rin=25 AU and log\macc=$-$9.8 is Ser34. }
        \label{fig::MaccvsRin}
\end{figure}

In Fig.~\ref{fig::MaccvsRin} we show the logarithmic values of \macc \ determined in Sect.~\ref{sect::method} as a function of the values of \rin \ reported in the literature (see Table~\ref{tab::XS}). We represent these values using different symbols to differentiate measurement of \rin \ derived with resolved mm-interferometry observations ({\it red circles}) from those obtained by modeling the optical to mid-infrared SEDs ({\it blue squares}). The uncertainties on the values of \rin \ are various and depend strongly on the assumptions on the models. In particular, values of \rin \ determined with SED fitting are strongly model-dependent and can be an overestimation of the real gap size \citep{Rodgers-Lee14}. 
We do not find any strong trend of \macc \ increasing with \rin \ over the whole range of \rin \ we have explored. If we compare our results with more complete samples of TDs, e.g., \citet{Kim13}, we see that our results agree with the upper boundary of their sample, and that indeed there is an increase of \macc \ with \rin \ up to values of \rin$\sim$ 20 AU. At \rin \ larger than about 20 AU, however, \macc \ is essentially constant in our sample. In fact, a similar trend is also present in the upper envelope of the TDs considered in \citet{Kim13}, where, however, there are just two accreting TDs with \macc$>10^{-8}M_\odot$/yr for \rin$\gtrsim$ 20 AU and none for \rin$\gtrsim$30-40 AU. All objects in our sample have \macc \ in the range $10^{-9}-10^{-8} M_\odot$/yr, independently of the value of \rin. Therefore, the density of their innermost gaseous disk, which accretes onto the star, does not depend on the mechanism that produces the gap or the hole, and must be high enough to sustain the observed accretion rates.

\begin{figure}
\centering
\includegraphics[width=0.5\textwidth]{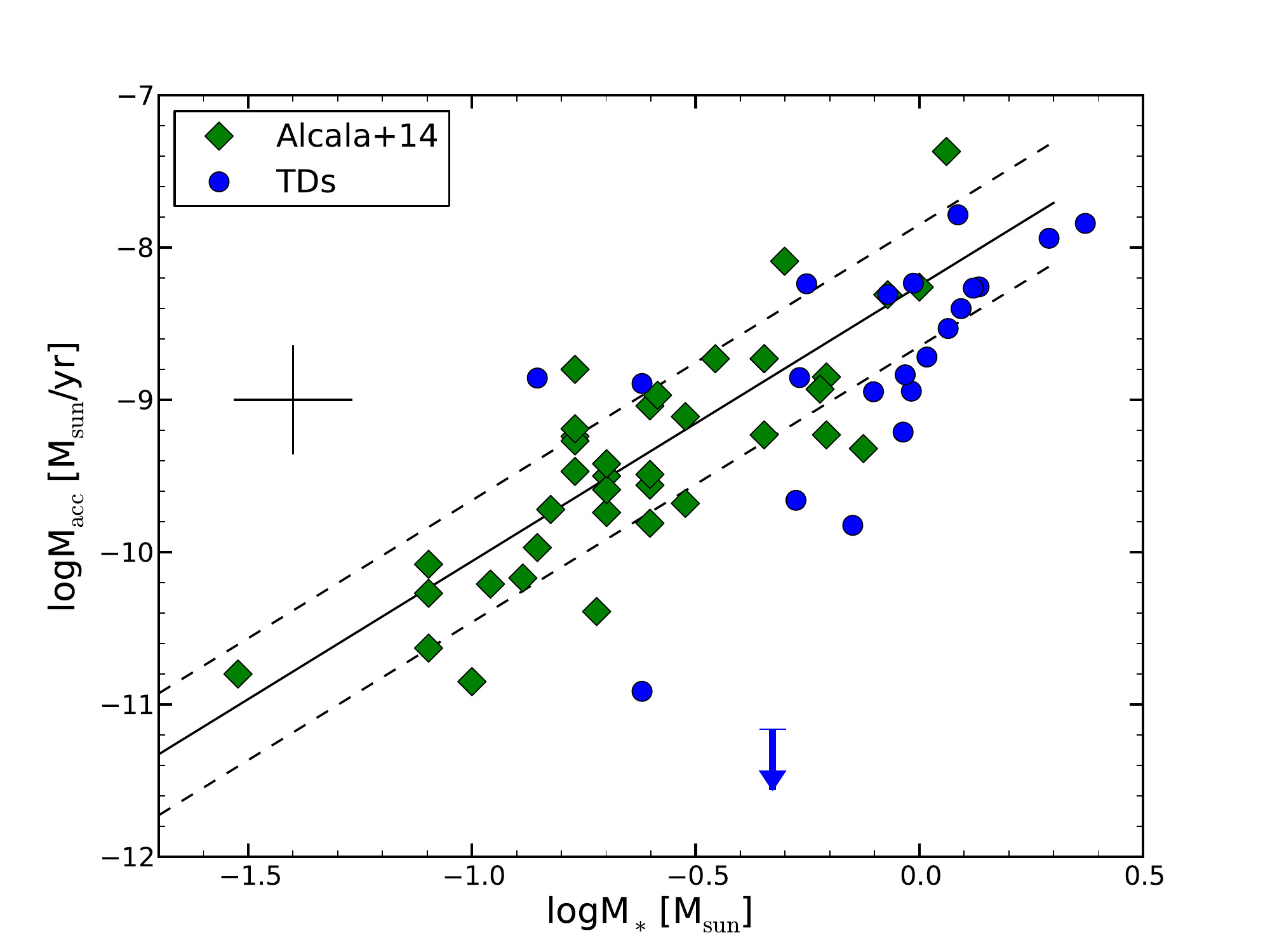}
\caption{Logarithm of the mass accretion rate vs logarithm of the stellar mass for our sample of TDs and for a sample of classical TTauri stars from \citet{Alcala14}. Our targets are shown as {\it blue circles}, while data from the literature are reported with {\it green diamonds}. The lines are the best fit to the data reported in \citet{Alcala14} (solid line) and the 0.4 dex spread reported in that study. Downward arrows are upper limits. The two lowest points are, from left to right, CHXR22E and Ser29. Typical errors are shown with the black cross.}
        \label{fig::MaccvsMstar}
\end{figure}

We want now to compare the derived values of \macc \ for our sample of TDs with a sample of classical TTauri stars (cTTs) to determine whether the accretion properties are different in these two classes of objects. It is well established that the values of \macc \ in cTTs depend on $M_*$ with a power of $\sim$1.6-1.8 \citep[e.g.,][]{Muzerolle03,Rigliaco11,Manara12,Alcala14,Ercolano14}. A comparison of the values of \macc \ between different samples should be based on a comparison of this relation and not on the values of \macc \ alone. Another well known dependence is the one between \macc \ and the age of the targets \citep[e.g.,][]{Hartmann98,Sicilia-Aguilar10,Manara12}, which is a consequence of the viscous evolution of protoplanetary disks. Therefore, a comparison should be carried out between objects of similar mean age. Finally, different methodology and evolutionary models can lead to different values of \macc; it is thus needed to compare samples analyzed with a similar methodology. For these reasons we select as a comparison sample the objects studied by \citet{Alcala14}. These are located in the Lupus I and III clouds and have ages $\sim$ 3 Myr, similar to the objects in our sample. The analysis of that sample was carried out with the same methodology as the one we used here. We show in Fig.~\ref{fig::MaccvsMstar} the logarithmic relation between \macc \ and $M_*$ for these two samples. Our data are reported as blue circles, while data from \citet{Alcala14} as green diamonds. The solid line on the plot is the best fit relation from \citet{Alcala14}, and the dashed lines represent the dispersion of 0.4 dex around this best fit relation. The typical errors on the quantities are shown as a black cross. We see that $\sim$80\% of the TDs have values of \macc \ consistent with the values found by \citet{Alcala14} for Lupus objects of the same $M_*$. Therefore, for these objects we do not see any difference in the accretion properties with respect to those in cTTs. We also perform on these two samples a Kolmogorov-Smirnof statistical test (K-S test). When restricting to objects in the same $M_*$ range the probability that the two samples are drawn from the same distribution is 80\%. 
We can then conclude that for our sample, the amount of accretion depends on the mass of the central object, and not on the evolutionary stage (cTTs or TD) of the system.

This result differs from what found in the literature. For example, \citet{Najita07} noted that TDs have a sistematically smaller value of \macc \ at any given value of the mass of the disk ($M_d$). They inferred that the accretion rates for TDs are in general smaller than for cTTs. Similarly, various further analyses of larger samples of TDs found values of \macc \ typically lower than those of cTTs by a factor $\sim$10 \citep[e.g.,][]{Kim09,Kim13,Muzerolle10,Espaillat12}. These results are reported in the recent review by \citet{EspaillatPPVI}, together with other results that differ somehow from the ones listed above. In particular, \citet{EspaillatPPVI} report values of \macc \ for 3 TDs in $\rho$-Ophiucus which are compatible with the locus of cTTs in the same region and in Taurus on the \macc-$M_d$ plane. They suggest that results differing from those found in previous works may arise from different sample selection and/or different methods to estimate \macc. To avoid this possible methodological bias, we have shown here only the comparison between our sample of TDs and the sample of cTTs in Lupus, which is analyzed in the same way as our objects. We stress here again that our TDs have been selected to be mostly strong accretors, thus our sample selection is not representative of all the TDs and that we do not derive conclusions for the whole TD population. Nevertheless, our results prove that there are TDs that accrete at the same rate of cTTs.

The two main outliers in Fig.~\ref{fig::MaccvsRin}-\ref{fig::MaccvsMstar} are the object with an upper limit on \lacc, Ser29, and CHXR22E. The lower intensity of accretion for these targets implies that the gas density in the inner disk is substantially depleted with respect to the one of cTTs. These objects do not have any peculiar property reported in the literature. From Fig.~\ref{fig::MaccvsMstar} we note that in the same range of $M_*$ of these objects there are other TDs with values of \macc \ comparable or even higher than cTTs. Therefore, these objects are not peculiar in their stellar properties. We will discuss in more detail about these objects later after considering their wind and dusty inner disk properties. It is possible that these objects are part of a population of TDs with lower values of \macc \ not included in our sample.

\subsection{Wind properties}

\begin{figure}
\includegraphics[width=0.5\textwidth]{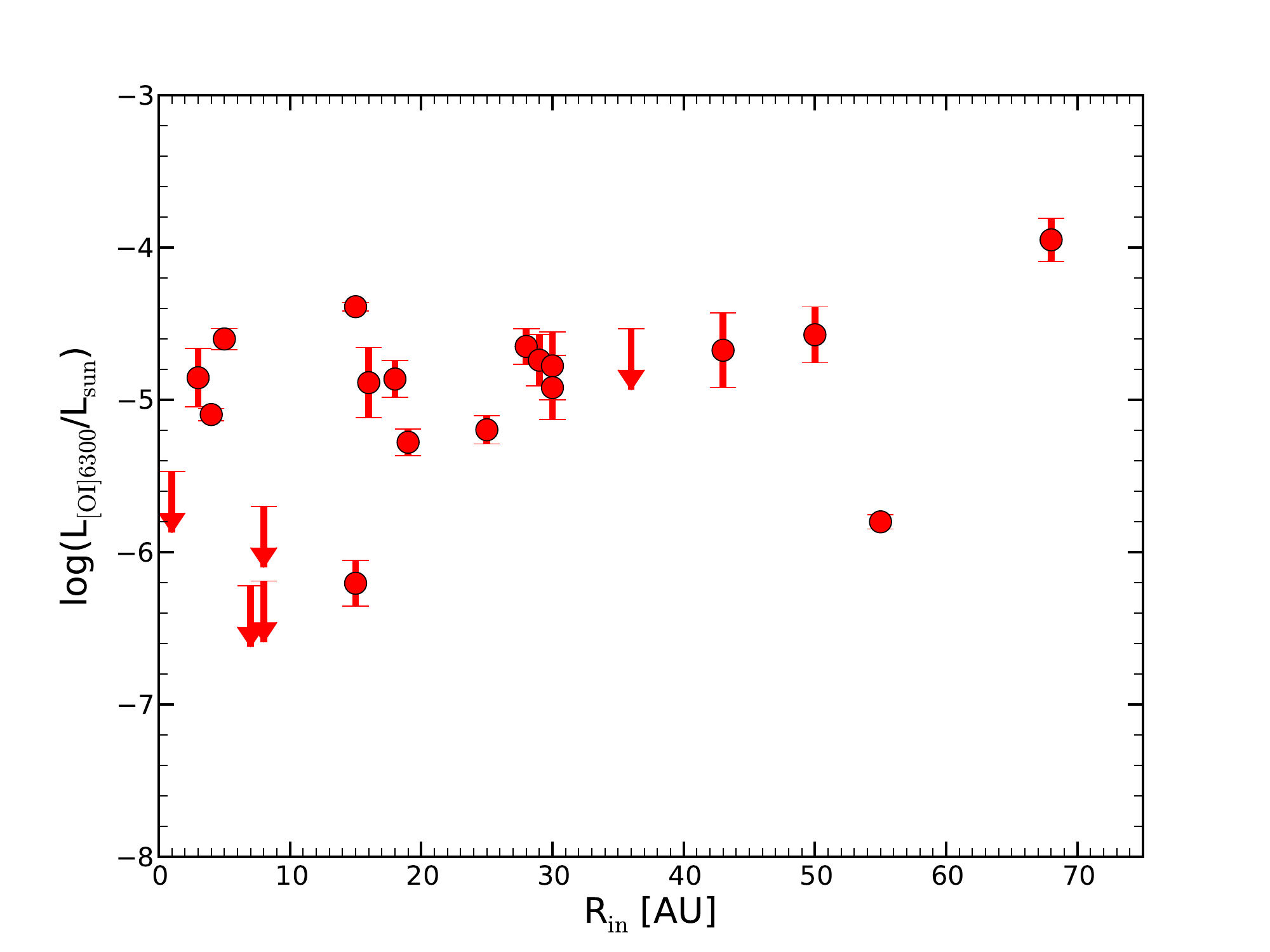}
\caption{Logarithmic luminosity of the low-velocity component of the [OI] 630 nm line vs inner hole size for the TDs in our sample. Downward arrows are upper limits. }
        \label{fig::LOI_Rin}
\end{figure}

\begin{figure}
\includegraphics[width=0.5\textwidth]{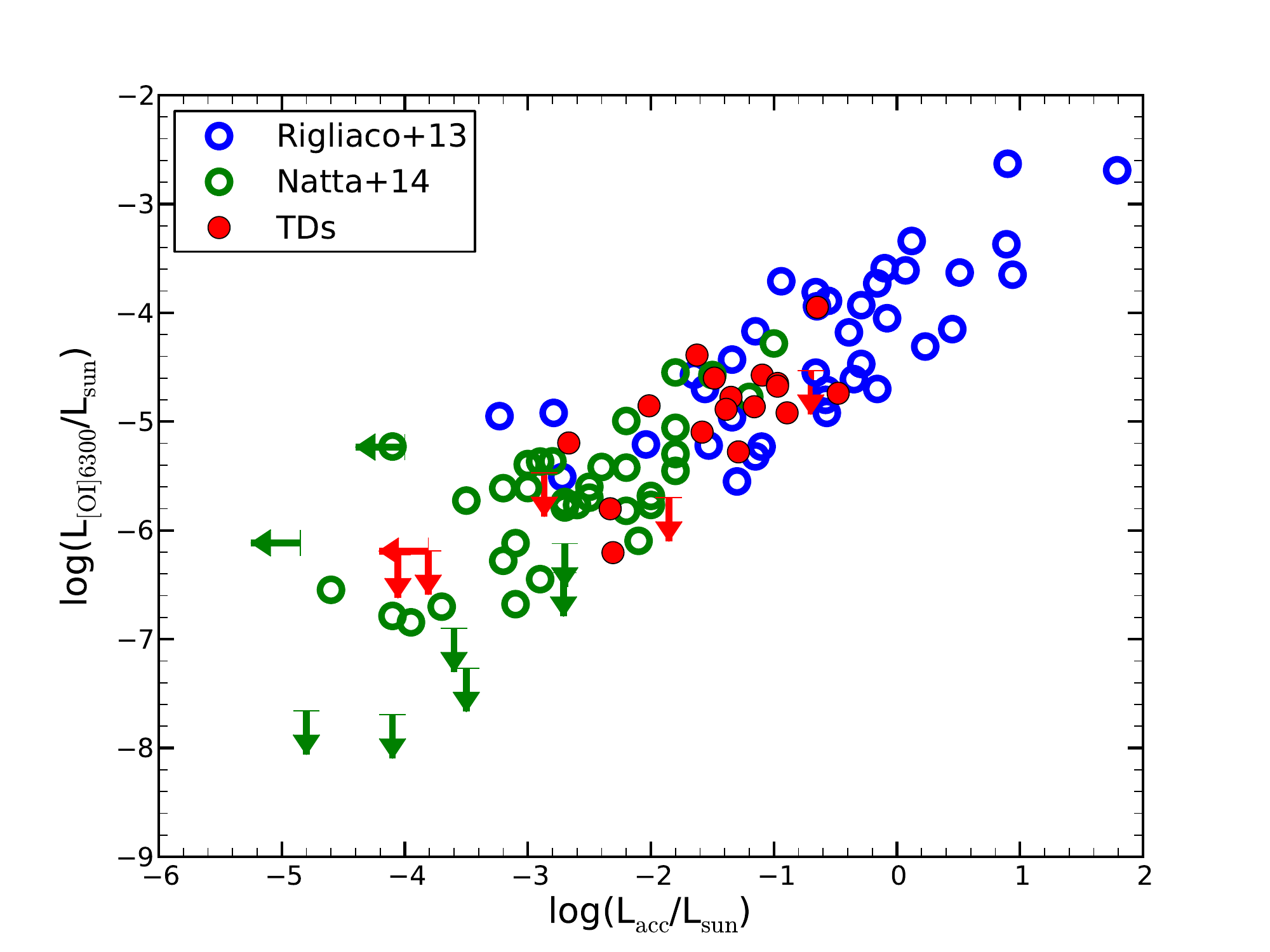}
\caption{Logarithmic luminosity of the low-velocity component of the [OI] 630 nm line vs the logarithm of the accretion luminosity of our objects ({\it red filled symbols}) and for two samples of classical TTauri stars ({\it blue empty circles} from \citealt{Rigliaco13}, {\it green empty symbols} from \citealt{Natta14}). Downward arrows are upper limits. }
        \label{fig::LOI}
\end{figure}

As discussed in Sect.~\ref{sect::wind}, we have measured the flux of the LVC of the [OI]$\lambda$630 nm line, which is a tracer of winds in YSOs. To determine whether the wind properties of our objects depend on the disk morphology we compare in Fig.~\ref{fig::LOI_Rin} the logarithmic luminosity of this line with the values of \rin \ available from the literature. We do not see any clear correlation between these quantities. The luminosity of the LVC of the [OI]$\lambda$630 nm line ($L_{\rm [OI]630}$) appears constant regardless the size of the dust depleted cavity in the disk with values between $\sim 10^{-6}$ and $\sim 10^{-4}$ \lsun. This implies that the properties of the wind traced by the [OI]$\lambda$630 nm line - that can be a disk wind, an accretion-driven wind, or a photoevaporative wind - are similar in most of the TDs in our sample. The question then is where in the disk the wind is originated. With the data in our hands we cannot put any constraint on the emitting region. Analysis of higher-resolution spectra of this line \citep[e.g.,][]{Rigliaco13} showed that, in cTTs, the emission region can be as close to the star as $\sim$ 0.2 AU, which is well within the dust depleted cavity in all our objects. Models of X-ray photoevaporation \citep{Ercolano10} predict that the luminosity of this line is insensitive to the size of the inner hole and depend mostly on the EUV and X-ray luminosity ($L_X$) of the central star. The X-ray photons are responsible for driving the wind in the first place, while the EUV photons heat up the inner region of the wind and excite the [OI] line. In this context, the correlation of \lacc \ with $L_{\rm [OI]}$ (see discussion in the next paragraph and Fig.~\ref{fig::LOI}) could be due to the heating of the wind by the UV photons. Therefore a lack of correlation of $L_{\rm [OI]}$ with $L_X$, that is found when comparing our data with the data reported in the literature for $L_X$ (see Table~\ref{tab::lit}), is not at all surprising, as the emission measure of this line is determined by the UV luminosity, which is instead correlated to \lacc. For this reason $L_{\rm [OI]}$ cannot be used as a quantitative tracer of the photoevaporated wind. Higher resolution spectra and more complete grids of models are needed to better constraint the origin of this line.

It is then important to compare the properties of this line in our TDs and in cTTs to understand whether they are similar in the two classes of objects. As comparison samples we select the objects studied by \citet[][and references therein]{Rigliaco13} and those observed with X-Shooter and analyzed by \citet{Natta14}. These two samples are representative of different stellar, accretion, and wind properties of cTTs. In particular, the sample of \citet[][and references therein]{Rigliaco13} comprises mostly strong accretors with low to intermediate stellar mass, while \citet{Natta14} has a sample of low- and very low-mass YSOs with lower accretion rates. Here we compare only the luminosity of the LVC of the [OI]$\lambda$630 nm line derived in our work and in the comparison samples. The best way to compare these values is to analyze the $L_{\rm [OI]630}$-\lacc \ relation, which is well characterized in the literature \citep[see e.g.,][and references therein]{Rigliaco13}. This is shown in Fig.~\ref{fig::LOI}, where we plot $\log L_{\rm [OI] 630}$ as a function of $\log$ \lacc \ for our sample of TDs ({\it red filled circles}) and for the two samples of cTTs ({\it blue empty circles} for data from \citealt{Rigliaco13} and {\it green empty circles} for those from \citealt{Natta14}). The relation between these two quantities spans over $\sim$ 7 orders of magnitude in both axes with a typical spread of $\sim$ 1 dex for the cTTs, and our objects follow it very well in all the cases. The location of our TDs right in the middle between the two comparison samples reflects the fact that their accretion rates are typical of $\sim$0.5-1 $M_\odot$ YSOs, i.e. smaller than those in the sample of \citet{Rigliaco13} and larger than low-mass YSOs. At the same time, this implies that their wind properties traced by the [OI]$\lambda$630 nm line scale with the accretion properties in the same fashion as in cTTs. Therefore, the process responsible for the formation of this line should be the same in objects surrounded by dust-rich disks and in TDs. 

\subsubsection{[NeII] from the literature}

To better understand the properties of the winds in our objects we include also data from the literature on the [NeII]$\lambda$12.8 $\mu$m, which is a well know tracer of disk wind. This line has been detected in emission in the mid-infrared spectra of protoplanetary disks using {\it Spitzer} \citep[e.g.][and reference therein]{Pascucci07, Gudel10, Espaillat13} or ground-based observations \citep[e.g.,][]{Pascucci09,Pascucci11,Sacco12}. This line is of interest to study the inner gaseous disk properties as it traces warm gas ($T\sim$ 5000 K) and the effects of extreme ultraviolet (EUV) and X-ray emission from the star on the disk \citep{Glassgold07}. High-resolution spectroscopic studies constrained the emitting region of this line within 20-40 AU from the central star \citep{Sacco12}. High resolution observations of TW Hya, in particular, showed that most ($\gtrsim$80\%) of the [NeII] emission arises from the region where the disk is still optically thick, but still within $\sim$10 AU from the central star \citep{Pascucci11}. 

Among the objects in our sample, 13 have been observed with MIR spectroscopy and all of them have a [NeII] line detection, as we report in Table~\ref{tab::NeII_lit}. In all these objects we detected also the [OI] $\lambda$ 630 nm line, with the only exception being SR21.

\begin{table}
\centering
\caption{\label{tab::NeII_lit}Properties of [NeII] $\lambda$12.8$\mu$m line from the literature}
\begin{tabular}{lccccl}
\hline\hline
Name & F$_{\rm [NeII]\_hires}$ & FWHM & $v_{0}$   & F$_{\rm [NeII]\_Spitzer}$ & Ref \\
\hline
   LkH$\alpha$330 & ... & ... & ... & $0.38\pm0.19$	&	G10 \\
        DM Tau & ... & ... & ... & $0.55$ & G10  \\
       LkCa 15 & $<0.5$ & ... & ... & $0.28\pm0.02$ & S12  \\
        GM Aur & ... & ... & ... & $1.2\pm0.06$ & G10 \\
       Sz Cha & ... & ... & ... & $1.62\pm0.20$ & E13 \\
       TW Hya & $3.8\pm0.3$ & 14.6 & -6.2 & $5.9\pm1.1$ & P09,G10 \\
       CS Cha & $2.3\pm0.2$ & 27 & -3.3 & $3.63\pm0.07$ & P09,E13 \\
      CHXR22E & ... & ... & ... & ... & ... \\
         Sz18 & ... & ... & ... & ... & ... \\
         Sz27 & ... & ... & ... & $0.63\pm0.07$ & E13 \\
         Sz45 & ... & ... & ... & ... & ... \\
         Sz84 & ... & ... & ... & ... & ... \\
     RX J1615 & $1.4\pm0.2$ & 20.5 & -7.5 & $2.76\pm0.46$ & S12 \\
        Oph22 & ... & ... & ... & ... & ... \\
       Oph24 & ... & ... & ... & ... & ... \\
         SR 21 & $0.5\pm0.1$ & 15.1 & -8.3 & $<3.0$ & S12,G10 \\
         ISO-Oph196 & ... & ... & ... & ... & ... \\
       DoAr 44 & $<0.3$ & ... & ... & $0.68\pm0.33$ & S12,G10 \\
       Ser29 & ... & ... & ... & ... & ... \\
       Ser34 & ... & ... & ... & ... & ... \\
       RX J1842.9 & $<0.2$ & ... & ... & $0.43\pm0.13$ & S12,G10 \\
   RX J1852.3 & ... & ... & ... & $0.72\pm0.04$ & 	G10 \\

\hline 
\end{tabular} 
\tablebib{P11: \citet{Pascucci11}; P09: \citet{Pascucci09}; S12: \citet{Sacco12}; E13: \citet{Espaillat13}; G10: \citet{Gudel10}} 
\tablefoot{Fluxes are reported in units of $10^{-14}$ erg s$^{-1}$ cm$^{-2}$; $v_0$ and FWHM in units of km/s. }
\end{table}

\subsection{Accretion and wind properties in objects with inner disk emission}

Following the analysis described in Sect.~\ref{sect::method}, we divide the sample in two classes of objects: we refer to objects with no near-infrared color excess as TD, while those with excess are referred to as PTD, as reported in Table~\ref{tab::XS}. The morphological difference between these two classes is the presence of warm dust in the inner region of the disk of PTD, which could be a small ring of dust at few tenths of AU from the star \citep[e.g.,][]{Benisty10, Espaillat10}. This difference in the inner disk morphology could be due to different evolutionary stages of these two classes of objects or to different dust depleting mechanisms. Here we compare the accretion and wind properties of the objects in these two classes present in our sample to verify whether we see any difference among these objects. 

\begin{figure}
\centering
\includegraphics[width=0.5\textwidth]{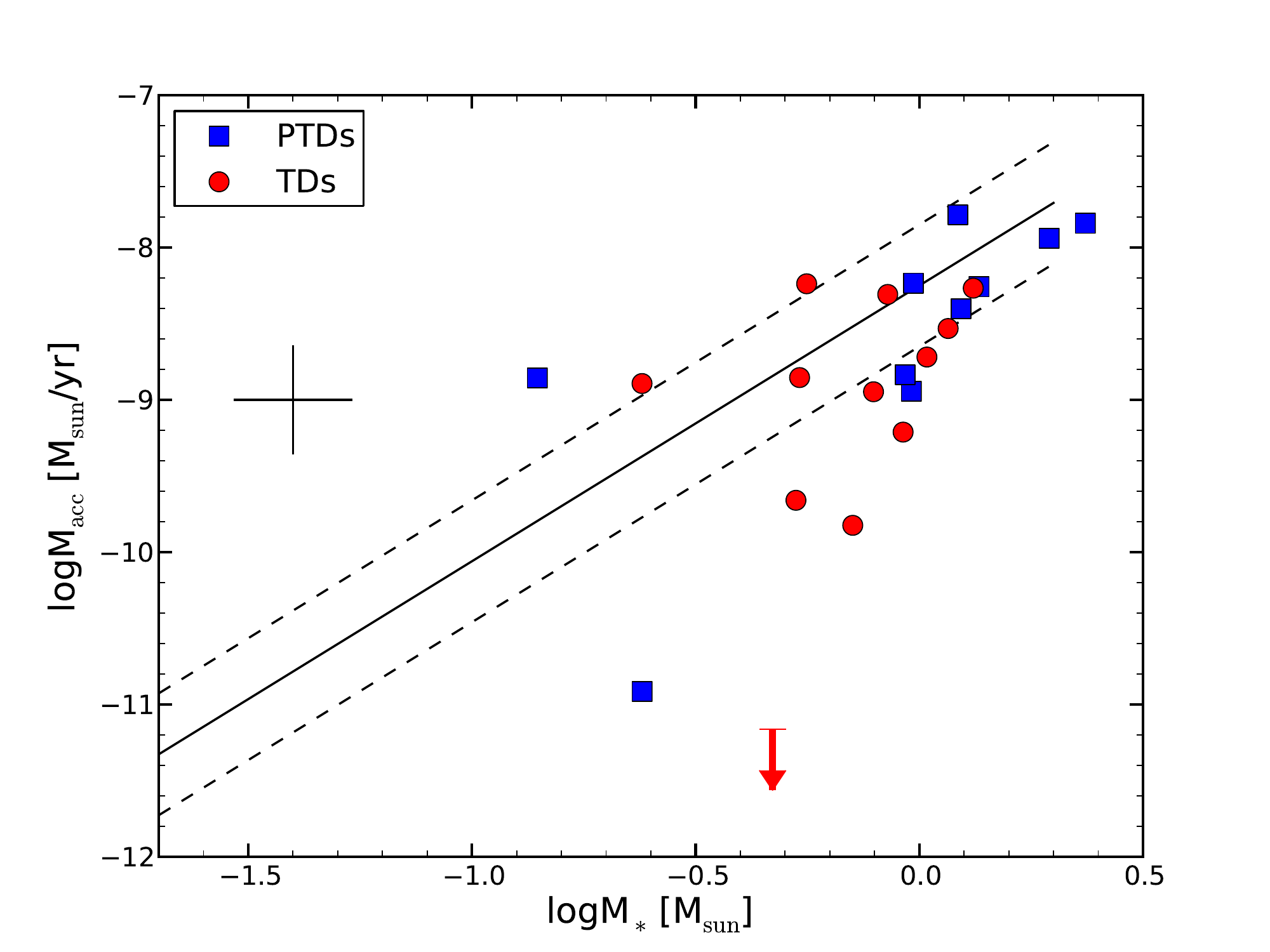}
\caption{Logarithm of the mass accretion rate vs logarithm of the stellar mass of our sample. Different symbols are used to distinguish objects with inner disk emission ({\it blue squares}) from TDs with no IR-excess ({\it red circles}). }
        \label{fig::MaccvsMstar_PTD}
\end{figure}

The comparison of the accretion properties of TDs and PTDs is carried out using the logartithmic \macc-$M_*$ relation, which is shown in Fig.~\ref{fig::MaccvsMstar_PTD}. In this figure different symbols are used to plot TDs ({\it red circles}) and PTDs ({\it blue squares}). We overplot here the best fit relation from \citet{Alcala14} as in Fig.~\ref{fig::MaccvsMstar} that is used as a reference. We clearly see that there is not a significant difference between the two classes of objects. Similarly to what we discuss in Sect.~\ref{sect::accretion}, also this result is apparently at odd with previous studies, which showed that PTDs accrete at a lower rate than TDs \citep[e.g.,][]{Espaillat10,Kim13}. This implies that the accretion properties in our sample of TDs are independent on the dusty inner disk morphology.

\begin{figure}
\includegraphics[width=0.5\textwidth]{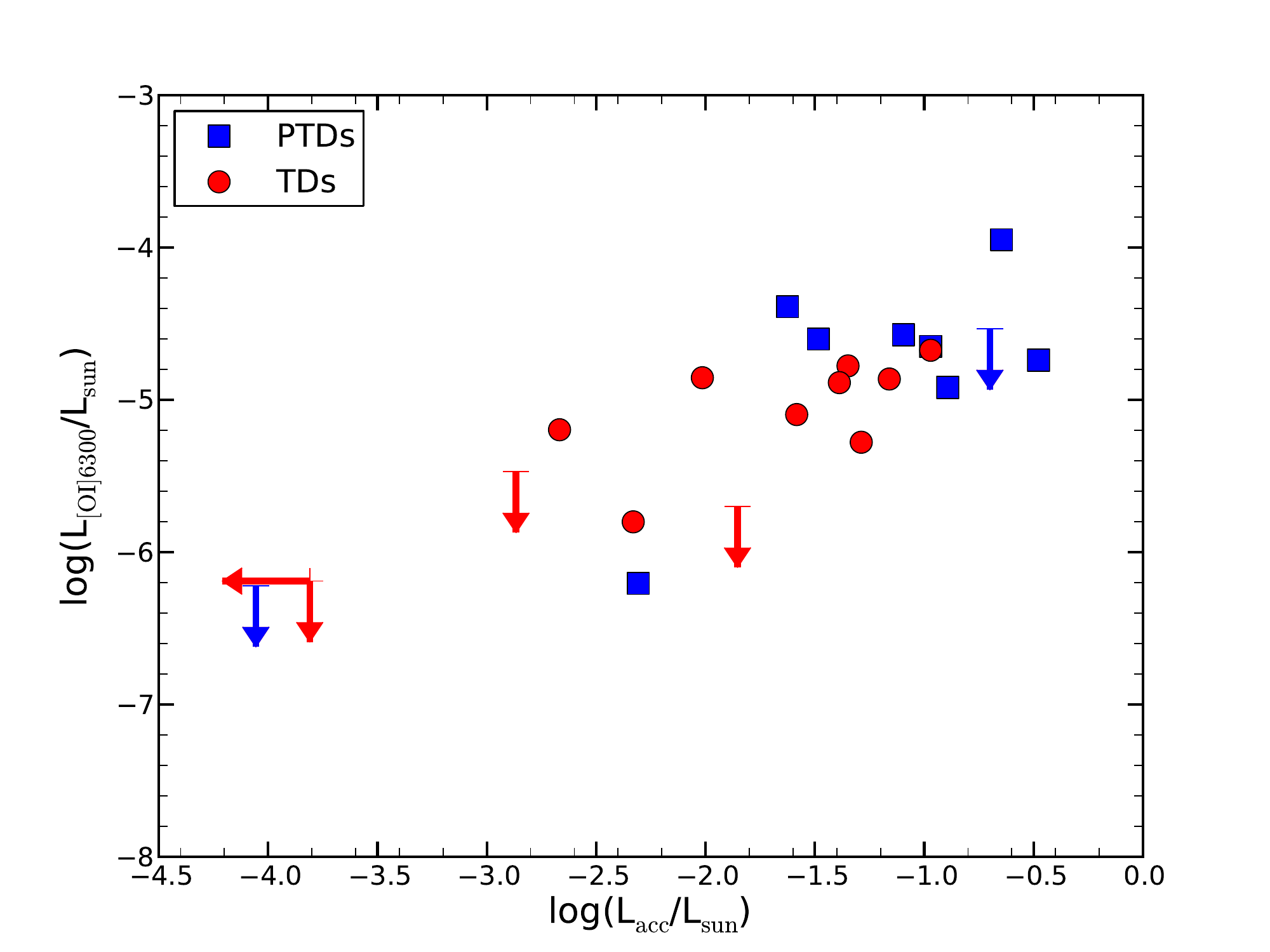}
\caption{Logarithmic luminosity of the low-velocity component of the [OI] 630 nm line vs the logarithm of the accretion luminosity of our objects. Symbols and colors are the same as in Fig.~\ref{fig::MaccvsMstar_PTD}. }
        \label{fig::LOI_PTD}
\end{figure}

We proceed with this analysis by comparing in Fig.~\ref{fig::LOI_PTD} the logarithmic relation between \lacc \ and $L_{\rm [OI] 630}$ for our sample, using different symbols for TDs ({\it red circles}) and for PTDs ({\it blue squares}) to see whether wind properties depend on the inner disk properties. Also in this case there is no correlation between the position on the plot and the inner disk morphology. The wind properties traced by the [OI] $\lambda$ 630 nm line are thus independent of the presence of dust in the innermost region of the disk.

\subsection{Constraint on the gas content of the inner disk}
\label{sect::constraint}

Here we present additional constraint on the region in the inner disk where gas is present and on its properties. From observations obtained in the literature we have measurement on the emission from CO from the inner part of the disk, which we discuss in the next subsection. Then, using the information on the accretion properties of our targets we derive the extent of the gas-rich inner disk, which extends down to the magnetospheric radius ($R_m$) at few stellar radii. We then also derive the density of gas in the inner disk needed to sustain the observed \macc \ if the disk is assumed in ``steady-state''. We also discuss possibilities to explain the observed \macc, and thus for a gas-rich inner disk, allowing for a significant gas depletion in the disk gaps. Finally, we give a complete view of the gas content of the inner disk adding to these informations also the wind properties of our TDs.

\begin{table}
\centering
\caption{\label{tab::CO_lit}Properties of CO fundamental transition from the literature.}
\begin{tabular}{lcl}
\hline\hline
Name & $R_{\rm in, CO}$ & Ref \\
\hbox{} & [AU] & \hbox{} \\
\hline
   LkH$\alpha$330	& 4$\pm$1 & P11		 \\
     DM Tau	& ...$^a$ & S09 \\
    LkCa 15	& 0.093	&	N03	 \\
     GM Aur 	& 0.5$\pm$0.2& S09 \\
    Sz Cha 	& ...$^c$ & ... \\
    TW Hya 	&   0.11$\pm$0.07	&	P08 \\
    CS Cha 	& ...& ... \\
   CHXR22E 	& ...& ... \\
      Sz18 	& ...& ... \\
      Sz27 	& ...& ... \\
      Sz45 	& ...& ... \\
      Sz84 	& ...& ... \\
  RX J1615 	& ...& ... \\
     Oph22 	& ...& ... \\
    Oph24 	& ...& ... \\
      SR 21 	& 7.6$\pm$0.4	& P08 \\
ISO-Oph196 	& ...& ... \\
    DoAr44 	& 0.4$\pm$0.1 & S09 \\
    Ser29 	& ...& ... \\
    Ser34 	& ...& ... \\
RX J1842.9 	& ...$^b$ & ... \\
RX J1852.3 	& ...$^c$ & ... \\

\hline 
\end{tabular} 
\tablebib{S09: \citet{Salyk09}; N03: \citet{Najita03}; P08: \citet{Pontoppidan08}; P11: \citet{Pontoppidan11}. $^a$ Non detection. Personal communications from A. Carmona: $^b$ detection of CO line, $^c$ non detection of CO line.} 
\end{table} 

\subsubsection{CO emission from the literature}
The fundamental ($\Delta v$ = 1) rovibrational line of CO at 4.7 $\mu$m is an important diagnostic to constraint the presence of gas in the inner region of TDs. This is sensitive to gas temperatures of 100-1000 K, which correspond to radii of 0.1-10 AU in typical protoplanetary disks around solar-mass stars YSOs, assuming that it originates in the so-called ``warm molecular layer" of the disk. Study of high-resolution spectra of this line have determined with high precision its emitting region within the disk \citep{Najita03,Salyk07,Pontoppidan08,Salyk09,Pontoppidan11}. These studies observed 7 objects included in this work, and detected the line in all of them besides DM Tau. We report in Table~\ref{tab::CO_lit} the inner radius of the CO emission in the disks ($R_{\rm in, CO}$) derived from the studies in the literature. Additional studies on three other objects of our sample have detected this line in RX J1842.9, while non-detection are obtained in the spectra of Sz Cha and RX J1852.3 (A. Carmona, personal communication). Seven out of the ten objects where this line has been studied are classified as PTD, so this emission could arise from the dusty inner disk. However, we also see that CO can be emitted in the dust-depleted inner disk of some TDs, such as TW Hya or RX J1852.3. A possible explanation for the emission of CO in the case of TW Hya is local warming due to the presence of a companion orbiting in the gap \citep{Arnold12}. In any case, the detection of CO emission in the aforementioned objects confirms that their inner disk is gas-rich, confirming the results obtained by detecting ongoing accretion in the same targets.

\subsubsection{Magnetospheric radius}
\label{sect::rm}

In the context of magnetospheric accretion models \citep[e.g.,][]{Hartmann98} the position in the disk from where the gas is accreted onto the star is determined by $R_m$. This is the radius where the external torque due to star-disk magnetic interaction dominates over the viscous torque. Following \citet{Armitage10}, this can be derived by equating the expressions representing the two timescales involved in this process, namely the magnetospheric accretion timescale:
\begin{equation}
t_m \sim 2\pi \frac{\Sigma \sqrt{GM_* r}}{B_z^s B_\phi^s r},
\end{equation}
where $r$ is the radial distance in the disk from the central star, $B$ is the magnetic field, the superscript $s$ stands for magnetic field evaluated at the disk surface, and $\Sigma$ is the surface density of the gas, and the viscous timescales:
\begin{equation}
t_\nu \sim \frac{r^2}{\nu},
\end{equation}
where $\nu$ is the disk viscosity. We assume the steady-state disk relation for the viscosity:
\begin{equation}
\nu\Sigma = \frac{\dot{M}_{\rm acc}}{3\pi},
\label{eq::st_st}
\end{equation}
that implies a constant \macc \ in the disk, and we consider the simple case where the stellar magnetic field is bipolar and oriented in the same direction as the rotation axis of the star. With these assumptions, we derive the usual relation for $R_m$ \citep[e.g.][]{Hartmann09}:
\begin{equation}
R_m \sim \left( \frac{3B_*^2R_*^6}{2\dot{M}_{\rm acc}\sqrt{GM_*}} \right)^{2/7}.
\end{equation}
It is important to note that this quantity depends weakly on \macc, $M_*$, and to $B_*$. The stronger dependence is on $R_*$. Using this relation we are able to derive the values of $R_m$ for all the accreting TDs in our sample using the values of \macc\footnote{The values of \macc \ derived previously have been obtained assuming $R_m$ = 5 $R_*$. This is the usual assumption made in the literature, thus this value is the appropriate one to derive \macc \ consistently  to previous analyses. The same values of \macc \ can be adopted here because of the weak dependence of $R_m$ on \macc \ ($R_m \propto \dot{M}_{\rm acc}^{-2/7}$). By re-deriving \macc \ using the newly determined $R_m$ we obtain values of \macc \ with a typical difference of $\sim$0.05 dex and always smaller than 0.1 dex. This translates in relative uncertainties on the value of $R_m$ of less than 0.06.}, $M_*$, and $R_*$ derived in Sect.~\ref{sect::method}, and assuming a typical value for the magnetic field of the star $B_* \sim 1$ kG \citep[e.g.][]{Johnstone13}.  The effect of the arbitrary choice of the value of $B_*$ is the prominent source of uncertainty in our estimate of $R_m$. We have to adopt a typical value for $B_*$ since only for two targets in our sample this quantity has been measured. This is the case of TW Hya and GM Aur, where $B_*$ is 1.76 kG and 1 kG, respectively \citep{Johns-Krull07}. By varying the values of $B_*$ from 2 kG to 0.5 kG we estimate a relative uncertainty on $R_m$ of less than 0.5. This is then the assumed uncertainty of our estimate.

The values of $R_m$ we have derived are reported in Table~\ref{tab::gas_prop}. In all the objects $R_m > 5 R_*$, in accordance with magnetospheric accretion models. This radius is always located at a distance from the central star much smaller than \rin. The detection of ongoing accretion implies that gas is present in the disk at this distance from the star. The gas density in the region of the disk at radii $\sim$ $R_m$ can be estimated as we explain in the next subsection.

\begin{table} 
\centering 
\caption{\label{tab::gas_prop}Derived properties of the gas} 
\begin{tabular}{lccc} 
\hline\hline 
Name & $R_m$  & $R_m$ & $\Sigma_{1AU}$  \\ 
\hbox{} & [$R_*$] & [AU]  & [g cm$^{-2}$] \\ 
\hline 
LkH$\alpha$330 & 10.05  & 0.175 & 411.36  \\ 
DM Tau & 7.94  & 0.058 & 106.87  \\ 
LkCa 15 & 8.35  & 0.059 & 82.81  \\ 
GM Aur & 7.46  & 0.061 & 174.28  \\ 
Sz Cha & 5.54  & 0.039 & 337.56  \\ 
TW Hya & 8.37  & 0.033 & 19.17  \\ 
CS Cha & 8.07  & 0.062 & 116.45  \\ 
CHXR22E & 35.18  & 0.134 & 0.12  \\ 
Sz18 & 10.88  & 0.068 & 23.91  \\ 
Sz27 & 9.73  & 0.052 & 24.61  \\ 
Sz45 & 7.50  & 0.053 & 118.54  \\ 
Sz84 & 14.55  & 0.113 & 15.01  \\ 
RX J1615 & 9.49  & 0.084 & 92.24  \\ 
Oph22 & 24.22  & 0.240 & 4.63  \\ 
Oph24 & 13.13  & 0.089 & 15.08  \\ 
SR 21 & 8.97  & 0.117 & 299.26  \\ 
ISO$-$Oph196 & 11.51  & 0.054 & 9.45  \\ 
DoAr 44 & 6.28  & 0.032 & 102.45  \\ 
Ser29 & ... & ... & ...  \\ 
Ser34 & 18.61  & 0.106 & 2.95  \\ 
RX J1842.9 & 8.88  & 0.043 & 25.22  \\ 
RX J1852.3 & 9.08  & 0.051 & 34.91  \\ 
\hline 
\end{tabular} 
\end{table}

\subsubsection{Density of gas in the inner disk}
\label{sect::density}

Assuming steady-state disk condition the surface density of the gas is related to the accretion disk viscosity and \macc \ by the relation reported in Eq.~(\ref{eq::st_st}). We describe the viscosity using the $\alpha$ viscosity prescription \citep[$\nu = \alpha c_s H$,][]{Shakura} and we assume that the disk is vertically isothermal, so that $H=c_s/\Omega(r)$, where $c_s = (k T/\mu m_p)^{1/2}$ is the sound speed, $\mu$=2.3 is the mean molecular weight, $m_p$ is the mass of the proton, and $\Omega = (GM_*/r^3)^{1/2}$ is the angular velocity of the disk. We then derive the following relation for the surface density of the gas in the disk:
\begin{equation}
\Sigma(r) \sim \frac{2 m_p}{3\pi \alpha k_B T(r)} \dot{M}_{\rm acc} \sqrt{\frac{GM_*}{r^3} } .
\end{equation}
We estimate the surface density of the gas at a distance of 1 AU from the central star ($\Sigma_{1 AU}$). This radius is chosen because it is much larger than $R_m$ but still within \rin \ for all our targets. Assuming $\alpha=10^{-2}$ and $T$(1AU) = 200 K \citep[representative value derived from][]{Andrews07}, we derive the values of $\Sigma_{1 AU}$ from the central star reported in Table~\ref{tab::gas_prop}. These values vary from few g cm$^{-2}$ to $\sim 4 \times 10^2$ g cm$^{-2}$ for our objects, and represent the expected densities of gas in the disk inner region needed to sustain the observed accretion rates assuming steady state viscous inner disk. Another possibility is that the density of the gas in the cavity is lower than the one derived here if the radial inflow of gas is at high velocity, approaching free-fall \citep{Rosenfeld14}. Finally, episodic events that replenish the gas content of the inner disk from the outer disk could also possibly explain our observed \macc \ with a significantly gas depleted hole for most of the TD lifetime.

\subsection{Discussion on the gas content of the inner disk}

We now want to put together all the information collected from our spectra and from the literature on the objects in our sample to understand what is the morphology of their gaseous inner disk. The discussion is divided between accreting and non-accreting objects. All the objects analyzed in this work besides Ser 29 have accretion detected with our method. Also CHXR22E has a measured value of \macc \ lower than other objects with similar stellar properties, as we pointed out when discussing the result of Fig.~\ref{fig::MaccvsMstar}. We discuss these two objects in Sect.~\ref{sect::nonacc}, while the other 20 accreting objects are discussed in the next subsection.

\subsubsection{Accreting transitional disks} 

As discussed in the introduction, the detection of measurable accretion rates in young stellar objects implies that the innermost region of the disk is gas rich. This is the case for our accreting TDs, and we derived in Sect.~\ref{sect::rm} and \ref{sect::density} the inner boundary of the gaseous disk in these objects and the densities of the gas in the inner disk needed to sustain the observed accretion rates assuming a steady-state viscous disk. We constrain with our analysis that gas is present in these disks in regions as close to the star as $\sim$ 0.03 - 0.3 AU, that are the values of $R_m$. The evidence of gas presence in this region is confirmed in 17 of the 20 accreting TDs with the detection of the [OI] $\lambda$ 630 nm line in their spectra, which is originated as close as $\sim$ 0.2 AU from the star. At similar disk radii ($\sim$ 0.1-0.5 AU) the CO emission is detected in 4 objects (LkCa15, GM Aur, TW Hya, and DoAr 44, see Table~\ref{tab::CO_lit}), confirming the presence of gas in their inner disk. For TW Hya this region is known to be strongly dust-depleted. On the other hand, it is plausible that the CO emission arises from the dusty inner disk in the other three objects, known to be PTDs. To these objects we should add RX J1842.9 where the CO line is also detected, but no analysis has yet been carried out to determine the distance to the star of the region emitting this line. For LkH$\alpha$330 and SR21 the emission of the CO line arises from larger radii ($R_{\rm CO} \ge$ 4 AU) due to the higher temperature of the disk related to the larger $L_*$ of these objects compared to the rest of the sample. Finally, in 13 of the 20 accreting TDs we could find a detection of the [NeII] in the literature. This line is also originated in a wind coming from a gas rich region of the disk inside a distance from the central star of $\sim$20-40 AU. In only 9 of the 20 accreting TDs we found evidence of infrared excess, a signature of the presence of a dusty inner disk. 

The picture of these accreting TDs coming from our analysis is then the following. These are objects with a gas rich disk well within the observed \rin, i.e. at the inner disk edge. Given that we do not find any correlation between the dust-depleted hole and the accretion or wind properties, and that we see both accreting TDs with dusty inner disks and without, the model needed to explain the formation of the dust-depleted inner region should leave almost unaltered the gas properties of the innermost region.  From the point of view of the gas content of the inner disk there is no observable difference between accreting TDs and cTTs.

\subsubsection{Non-accreting transitional disks} 
\label{sect::nonacc}

The two objects we discuss here (Ser 29 and CHXR22E) have a gas-depleted inner region of the disk. The non-detection of accretion signatures in these objects or the very low detected \macc \ of CHXR22E imply that the density of the gas in this region of the disk is smaller in these objects by at least one order of magnitude than in any other accreting object. This is clearly seen in the value of $\Sigma_{\rm 1 AU}$ = 0.1 g cm$^{-2}$ reported for CHXR22E in Table~\ref{tab::gas_prop}, smaller by a factor $\sim$ 30-40 than the one computed for Oph 22 and Ser 34, that have similar stellar properties. With the addition of the non-detection of the [OI] $\lambda$ 630 nm in the spectra of both objects we conclude that the region around $\sim$ 0.1-0.3 AU is significantly gas-depleted in these non-accreting TDs. No further information on the gas content of the inner disk of Ser 29 and CHXR22E are available. These two targets should be observed in the future with the aim of detecting [NeII] and/or CO emission in the inner parts of these objects, in order to constraint the inner boundary of the gas-rich disk.


\section{Conclusions}
\label{sect::concl}
In this work we analyzed a sample of 22 X-Shooter spectra of TDs. This sample comprises objects with different outer disk morphologies, in particular with values of \rin \ ranging from $\sim$1 AU to $\sim$ 70 AU, and includes mainly TDs with previous accretion rate estimates. This sample cannot provide a conclusive statistical result on the general properties of the TD class, but it is a good benchmark to study with an highly reliable method these objects. We used a multi-component fitting method to derive simultaneously the SpT, $A_V$, and \lacc \ of the objects fitting our broad-band spectra. At the same time we derived from the same spectra the intensity of the [OI]$\lambda$ 630 nm line. From the analysis of the results we derived the following conclusions:

- The dependence of the accretion properties of our sample of strongly accreting TDs with the size of the dust-depleted cavity (\rin) is small, in particular there is no evidence for increasing \macc \ with \rin \ at values of \rin$\gtrsim 20-30$ AU;

- There are strongly accreting TDs, like the majority of the objects in our sample, whose accretion properties are consistent with those reported in the literature for cTTs;

- The wind properties of the TDs analyzed here have no dependence with the size of the dust-depleted cavity (\rin) and are consistent with the wind properties of cTTs;

- There are no differences in the accretion and wind properties between the objects in our sample with inner disk emission (PTD) or without (TD);

- Strongly accreting TDs such as those analyzed here are gas-rich down to distances from the central star as small as $\sim$0.03-0.3 AU as can be obtained from the derivation of the values of $R_m$, from the detection of the [OI] $\lambda$ 630 nm line, and from the detection of the CO and [NeII] lines. This distance is always smaller than the values of \rin \ reported in the literature for these objects, meaning that there is a gaseous inner disk much closer to the star than the dusty one;

- Non-accreting TDs have gas depleted inner disks. The gaseous disk is significantly depleted of gas at a distance from the star of at least $\sim$0.03-0.3 AU. Also for these objects the inner extent of gas and dust in the disk are uncoupled;

- The process needed to explain the formation of TDs should act differently on the gas and the dust components of the disk.

Future studies aimed at understanding the process responsible for the formation of the dust depleted cavity in TDs should aim at:

- conducting a similar analysis on a larger and more complete sample of TDs, including a larger amount of objects known to accrete at lower rates than those included in this work;

- determining the process responsible for the formation of the forbidden lines - photoevaporation, disk wind, or other possibilities - from the analysis of high-resolution and high-signal to noise forbidden lines and the comparison with theoretical models covering more completely the parameter space of stellar properties;

- determining the extent of the region emitting the [OI] and [NeII] line and the density of gas in this region in order to put stronger constraint on the distance from the central star at which gas is present;

- studying the CO line with high-resolution spectroscopy in non-accreting TDs to verify the decoupling of the gas-rich and dust-rich disk in these objects.

\begin{acknowledgements}
We thank the ESO staff in Paranal for carrying out the observations in Service mode. We thank J. Alcala and the ``JEts and Disk at Inaf'' (JEDI) team for providing the reduced spectrum of Sz84. We thank A. Carmona for sharing the information on the CO line detection. C.F.M. acknowledges the PhD fellowship of the International Max-Planck-Research School. 
\end{acknowledgements}

\appendix

\section{Comments on individual objects}

\subsection{Sample properties}
\label{app::peculiar}
{\bf LkCa 15}: this object has been resolved with 880 $\mu$m interferometric observations by \citet{Andrews11}, and its cavity was previously resolved at 1.3 mm by \citet{Pietu06}. The modelling of this target carried out by \citet{Andrews11} has a discrepancy with the observed 880 $\mu$m flux inside the cavity probably due to dust emission.\\
{\bf Sz Cha}: we observed only the primary component of this wide binary system with separation 5\farcs122. The companion of this object is not a confirmed member of the Cha~I association \citep{Luhman08}.\\
{\bf CS Cha}: \citet{Guenther07} classified this objects as a spectroscopic binary with a period of more than 2482 days. The minimum mass of the companion is 0.1 \msun. \\
{\bf Sz 84}: It is under debate whether this object should be classified as TD. \citet{Merin10}, who classified it as a TD in first place, derived $R_{\rm in}$ = 55$\pm$5 AU, but they pointed out that the classification was rather uncertain due to possible extended emission contamination. Also \citet{Matra12} suggest that this classification is dubious. They point out that this object has no 10 $\mu$m silicate feature in the spectrum, which is a typical feature in the spectra of TDs. Then, they report that it has a SED very similar to the one of T54, which they propose is not a TD due to extended emission in the {\it Herschel} images.\\
{\bf RX J1615-3255}: different distances for this object are reported in the literature. \citet{Merin10} consider this object to be located in the $\rho$-Ophiuci cloud, thus at $d$=125 pc. \citet{Andrews11}, instead, adopt a distance to this object of 185 pc because they assume it is located in the Lupus cloud. This location is then adopted also by \citet{Sacco12}, but they use a distance of 150 pc for the object. We decide to adopt the distance of 185 pc used by \citet{Andrews11} for consistency with the values of \rin \ derived in that work. We then correct the value of \rin \ derived in \citet{Merin10} to the distance adopted here. This is the value of $R_{\rm in,SED}$ reported in Table~\ref{tab::lit}. \\
{\bf SR21}: this object is known to be a wide binary with a separation of $\sim$ 6\farcs4 $\sim$ 770 AU. We observed only the primary component of the system, which is the one observed by \citet{Andrews11}. Regarding the dust emission of this object, \citet{Andrews11} report a poor matching of the observations which may indicate that a small amount of $\sim$mm-sized dust particles is present in the cavity.\\
{\bf ISO-Oph196}: the inner dust depleted cavity has been barely resolved with SMA by \citet{Andrews11}. Looking at the SED of this object there is no signature of dust depletion, i.e. there is no dip in the MIR SED. This suggests that the dust is not strongly depleted in the inner disk of this object.

\section{Additional literature data}

We report in Table~\ref{tab::lit} additional data collected in the literature on our targets. These data have not been used in this work but are useful for further analysis. The spectral type and \macc \ reported here have not been used in our analysis.

\begin{table*} 
\centering 
\caption{\label{tab::lit}Stellar and disk parameters available in the literature} 
\begin{tabular}{lccccccccl} 
\hline\hline 
Name  & Spectral & \macc & $R_{\rm gap,in,SED}$ & $R_{\rm in,SED}$ & $R_{\rm in,mm}$ & $i$ & log$L_X$ & Disk & Ref \\ 
\hbox{}  & type & [10$^{-8}$ \msun/yr] & [AU] & [AU] & [AU] & [$\degree$] & [erg/s] & Type & \hbox{}\\ 
\hline 
LkH$\alpha$330  & G3 & 0.20 & 0.8 & 50 & 68 & 35 & ... & ... & 1,12,22,26 \\
DM Tau  & M1 & 0.60 & ... & 3 & 19 & 35 & 30.30 & TD & 1,10,20,21,28 \\
LkCa15  & K3 & 0.30 & 4 & 48 & 50 & 49 & $<$29.6 & PTD & 1,15,16,21,24 \\ 
GM Aur  & K5 & 1.00 & 1 & 24 & 28 & 55 & 30.20 & TD & 6,9,10,20,21,28 \\
SZ Cha  & K0 & 0.24 & ... & 29 & ... & ... & 29.90 & PTD & 16,17,21,25 \\ 
TW Hya & K7 & 0.20 & ... & 4 & 4 & 4 & 30.32 & ... & 10,14,23,27 \\ 
CS Cha  & K6 & 0.53 & ... & 43 & ... & 45 & 30.56 & TD & 10,16,24,25,30 \\ 
CHXR22E  & M3.5 & ... & ... & 7 & ... & ... & 29.41 & TD & 16,19  \\ 
Sz18  & M3 & 1.5e-10 & ... & 8 & ... & ... & ... & TD & 16,19 \\ 
Sz27  & M0 & 1.2e-9 & ... & 15 & ... & ... & 29.76$^{a}$ & PTD & 16,19  \\ 
Sz45 & M0.5 & 7.6e-10 & ... & 18 & ... & ... & ... & WTD & 16,19  \\ 
Sz84  & M5.5 & 1.00 & ... & 55 & ... & ... & ... & ... & 8 \\ 
RX J1615  & K5 & 0.04 & ... & 3$^{b}$ & 30 & 41 & 30.40 & ... & 1,2,8 \\ 
Oph22  & M2 & ... & ... & 1 & ... & ... & ... & ... & 8 \\ 
Oph24  & M0.5 & 1.00 & ... & 3 & ... & ...& ...  & ... & 8 \\  
SR 21  & G3 & $<$0.1 & 0.45 & 18 & 36 & 22 & 30.00 & ... & 1,3,4,5,18,26 \\ 
ISO-Oph 196  & M4 & 0.20 & ... & ... & 15 & 28 & ... & ... & 1,18 \\ 
DoAr 44  & K3 & 0.90 & ... & 27 & 30 & 35 & 29.9 & PTD & 1,13,29 \\ 
Ser29  & M0 & 30.00 & ... & 8 & ... & ...& ...  & ... & 8 \\ 
Ser34 & M0 & 0.25 & ... & 25 & ... & ...& ...  & ... & 8 \\ 
RX J1842.9  & K2 & 0.10 & ... & 5 & ... & ... & 30.34 & ... & 6,7,19 \\ 
RX J1852.3  & K3 & 0.05 & ... & 16 & ... & ... & 30.41 & ... & 6,7,19 \\ 
\hline 
\end{tabular} 
\tablebib{(1)~\citet{Andrews11}, (2)~\citet{Krautter97}, (3)~\citet{Brown09}, (4)~\citet{Andrews09}, (5)~\citet{Grosso00}, (6) ~\citet{Hughes10}, (7)~\citet{Neuhauser00}, (8)~\citet{Merin10}, (9)~\citet{Hughes08}, (10)~\citet{Gudel10}, (11)~\citet{Konig01}, (12)~\citet{Brown08}, (13)~\citet{Montmerle83}, (14)~\citet{Hughes07}, (15)~\citet{Neuhauser95}, (16)~\citet{Kim09}, (17)~\citet{White00}, (18)~\citet{Natta06}, (19)~\citet{Pascucci07}, (20)~\citet{Ingleby09}, (21)~\citet{Espaillat10}, (22)~\citet{Salyk09}, (23)~\citet{Herczeg08}, (24)~\citet{Ingleby13}, (25)~\citet{Espaillat13}, (26)~\citet{Brown07}, (27)~\citet{Calvet02}, (28)~\citet{Calvet05}, (29)~\citet{Kim13}, (30)~\citet{Pascucci09} ----- $R_{\rm gap,in,SED}$ is the inner radius of the gap in PTDs obtained from MIR SED fitting, whereas $R_{\rm in}$ is the inner radius of the dusty outer disk, i.e. the outer radius of the gap in PTDs and of the hole in TDs. $^{a}$ Highly uncertain parameter. $^{b}$ Value corrected for the different distance as explained in Appendix~\ref{app::peculiar}. } 
\end{table*}

\end{document}